\documentclass[aps,pre,twocolumn,superscriptaddress,showpacs]{revtex4}
\usepackage{amssymb,amsmath,latexsym,amsbsy,bm}
\usepackage{epsfig}
\usepackage{slashbox}
\usepackage{booktabs}
\usepackage[utf8]{inputenc}
\usepackage[T1]{fontenc}
\usepackage{graphicx}

\begin{document}

\title{Minimum spanning tree filtering of correlations for varying time scales and size of fluctuations}

\author{Jarosław Kwapień}
\author{Paweł Oświęcimka}
\author{Marcin Forczek}
\affiliation{Institute of Nuclear Physics, Polish Academy of Sciences, Kraków, Poland}
\author{Stanisław Drożdż}
\affiliation{Institute of Nuclear Physics, Polish Academy of Sciences, Kraków, Poland}
\affiliation{Faculty of Physics, Mathematics and Computer Science, Cracow University of Technology, Kraków, Poland}

\date{\today}

\begin{abstract}

Based on a recently proposed $q$-dependent detrended cross-correlation coefficient $\rho_q$ (J.~Kwapień, P.~Oświęcimka, S.~Drożdż, Phys. Rev.~E 92, 052815 (2015)), we generalize the concept of minimum spanning tree (MST) by introducing a family of $q$-dependent minimum spanning trees ($q$MST) that are selective to cross-correlations between different fluctuation amplitudes and different time scales of multivariate data. They inherit this ability directly from the coefficients $\rho_q$ that are processed here to construct a distance matrix being the input to the MST-constructing Kruskal's algorithm. Conventional MST with detrending corresponds in this context to $q=2$. In order to illustrate their performance, we apply the $q$MSTs to sample empirical data from the American stock market and discuss the results. We show that the $q$MST graphs can complement $\rho_q$ in disentangling ``hidden'' correlations that cannot be observed by the MST graphs based on $\rho_{\rm DCCA}$ and, therefore, they can be useful in many areas where the multivariate cross-correlations are of interest. As an example, we apply this method to empirical data from the stock market and show that by constructing the $q$MSTs for a spectrum of $q$ values we obtain more information about correlation structure of the data than by using $q=2$ only. More specifically, we show that two sets of signals that differ from each other statistically can give comparable trees for $q=2$, while only by using the trees for $q \ne 2$ we become able to distinguish between these sets. We also show that a family of $q$MSTs for a range of $q$ express the diversity of correlations in a manner resembling the multifractal analysis, where one computes a spectrum of the generalized fractal dimensions, the generalized Hurst exponents, or the multifractal singularity spectra: the more diverse the correlations are, the more variable the tree topology is for different $q$s. As regards the correlation structure of the stock market, our analysis exhibits that the stocks belonging to the same or similar industrial sectors are correlated via the fluctuations of moderate amplitudes, while the largest fluctuations often happen to synchronize in those stocks that do not neccessarily belong to the same industry.

\end{abstract}


\maketitle

\section{Introduction}
\label{sect.1}

Minimum spanning tree (MST) is a subgraph of a weighted network that minimizes sum of the edge lengths while spanning the network and not containing any cycles. It is often used in optimization and in multivariate data analysis to visualize the key properties of a network representation of the data if the corresponding complete network could not be shown in a transparent way. Many examples of the MST applications can be found in literature (see, e.g.,~\cite{barrow1985,adami1999,mantegna1999,zivkovic2006,hernandez2007}), but they are particularly common in econophysics~\cite{mantegna1999,vandewalle2001,onnela2003,micciche2003,bonanno2004,mcdonald2005,mizuno2006,eom2007,naylor2007,%
coelho2007,gorski2008,garas2008,sieczka2009,kwapien2009,smith2009,eryigit2009,eom2009,aste2010,keskin2011,zhang2011,%
sandoval2012,kwapien2012,zheng2013,wilinski2013,wang2013a,kantar2014,skowron2015}. Frequently, a network analysis is used to describe a correlation structure among a number of observables measured simultaneously. Such observables are represented by nodes and correlations by the weighted edges. These weights can be expressed by any correlation measure, like, for example, the Pearson coefficient or mutual information. Minimum spanning tree can then be built by using the Kruskal's~\cite{kruskal1956} or the Prim's algorithm~\cite{prim1957} provided the weights have been transformed into Euclidean distances.

One of the key problems that arise while dealing with empirical data is how to overcome nonstationarity. Standard correlation measures, like the already mentioned ones, are highly prone to instability caused by large fluctuations of the data~\cite{drozdz2001} and therefore results obtained with such measures are not fully reliable. Inevitably, the same weakness is inherited by the networks built on top of those measures. However, as the nonstationarities are typically associated with trends, simultaneous detrending on different scales can produce stationary data. The detrended fluctuation analysis (DFA) is the most frequently used method in this context~\cite{peng1994}. Its outcome, a fluctuation function, plays the role of variance for the nonstationary signals. Analogously, the role of covariance is played by the outcome of the detrended cross-correlation analysis (DCCA)~\cite{podobnik2008}. Based on both methods, the so-called detrended cross-correlation coefficient $\rho_{\rm DCCA}$ was introduced that is a counterpart of the Pearson correlation coefficient~\cite{zebende2011}. It can be used in the case of nonstationary data to quantify strength of cross-correlations between detrended signals at a given time scale~\cite{zebende2011,vassoler2012,zebende2013,reboredo2014,dasilva2015,hussain2017}.

In statistical analysis, variance and covariance are not able to provide a complete description of the probability distribution function for a given data set and the whole family of statistical moments is necessary to accomplish this. The Pearson correlation coefficient is not sensitive to any nonlinear dependences among data. In the same manner, DFA and DCCA are used to detect power-law auto- and cross-correlations and the related fractal properties of signals, but it is impossible to use them to detect nonlinear structures that are more complex than monofractals. Therefore, the detrended cross-correlation coefficient $\rho_{\rm DCCA}$ also has a limited power as regards the detection of higher-order statistics than a simple detrended covariance. In order to make it be more potent, recently we have porposed its generalization called the $q$-dependent detrended cross-correlation coefficient $\rho_q$ ($q \in \mathcal{R}$)~\cite{kwapien2015}. Its main purpose is to identify the cross-correlations selectively with respect to the fluctuation amplitudes by amplifying the correlated fluctuations of a particular size, while suppressing the ones of other size. The $\rho_q$ coefficient is based on the multiscale generalizations of DFA and DCCA, i.e., the multifractal detrended fluctuation analysis (MFDFA)~\cite{kantelhardt2002} and the multifractal detrended cross-correlation analysis (MFCCA)~\cite{oswiecimka2014}. Both these methods are principally applied to quantify multifractal structures in one or two signals, but their sensitivity to time scales makes them useful to describe also the signals that are not fractal in general.

The analogy between the $\rho_{\rm DCCA}$ coefficient and the Pearson coefficient can be exploited straightforwardly to construct the detrended minimum spanning trees~\cite{wang2013a}. Their main advantage over the standard MSTs is that they are more stable with respect to time as being independent of the data nonstationarity. In our present work we want to combine this property of the detrended minimum spanning trees with the ability of $\rho_q$ to focus on the fluctuations of specific size. As a result, we obtain a tool that is capable of quantifying correlation structure of multivariate nonstationary signals not only with respect to time scale but also with respect to fluctuation size. As in the case of $\rho_{\rm DCCA}$, the signals under study may be arbitrary, as no fractal properties are required.

Our paper is organized as follows. In Section~\ref{sect.2} we provide the necessary definitions, in Section~\ref{sect.3} we describe the empirical data and present results of the $q$MST analysis based on that data, and in Section~\ref{sect.4} we collect the main conclusions.

\section{Methods}
\label{sect.2}

Let us start from a brief description of the MFCCA approach~\cite{oswiecimka2014}. Consider a pair of time series ${x(i)}_{i=1,...,T}$ and ${y(i)}_{i=1,...,T}$ divided into $2 M_s$ boxes of length $s$ (i.e., $M_s$ boxes starting from the opposite ends). In each box $\nu$ ($\nu=0,...,2 M_s - 1$), we calculate the difference between the integrated signals and the $m$th-order polynomials $P^{(m)}$ fitted to these signals:
\begin{eqnarray}
X_{\nu}(s,i) = \sum_{j=1}^i x(\nu s + j) - P_{X,s,\nu}^{(m)}(j),\\
Y_{\nu}(s,i) = \sum_{j=1}^i y(\nu s + j) - P_{Y,s,\nu}^{(m)}(j),
\end{eqnarray}
where $m=2$ typically. The covariance and the variances of $X$ and $Y$ in a box $\nu$ are defined as:
\begin{eqnarray}
\label{eq::covariance}
f_{XY}^2(s,\nu) = {1 \over s} \sum_{i=1}^s X_{\nu}(s,i) Y_{\nu}(s,i),\\
f_{ZZ}^2(s,\nu) = {1 \over s} \sum_{i=1}^s Z_{\nu}^2(s,i).
\label{eq::variance}
\end{eqnarray}
Here $Z$ can be either $X$ or $Y$. Now we define a family of the fluctuation functions of order $q$~\cite{kantelhardt2002,oswiecimka2014}:
\begin{eqnarray}
\label{eq::covariance.q}
F_{XY}^q(s) = {1 \over 2 M_s} \sum_{\nu=0}^{2 M_s - 1} {\rm sign} \left[f_{XY}^2(s,\nu)\right] |f_{XY}^2(s,\nu)|^{q/2},\\
F_{ZZ}^q(s) = {1 \over 2 M_s} \sum_{\nu=0}^{2 M_s - 1} \left[f_{ZZ}^2(s,\nu)\right]^{q/2}.
\label{eq::variance.q}
\end{eqnarray}
The signum function enters the equation for $F_{XY}^q(s)$ in order to preserve signs of the covariances $f_{XY}^2(s,\nu)$ that would otherwise be lost in moduli that are necessary to secure real values of $F_{XY}^q(s)$. The real parameter $q$ plays the role of a filter, by amplifying or suppressing the intra-box variances and covariances in such a way that for $q \gg 2$ only the boxes (of size $s$) with the highest fluctuations contribute substantially to the sums while for $q \ll 2$ only the boxes with the smallest fluctuations do that. The special case of $q=2$ allows us to reduce the above formulas to a form:
\begin{eqnarray}
\label{eq::covariance.average}
F_{XY}^2(s) = {1 \over 2 M_s} \sum_{\nu=0}^{2 M_s - 1} f_{XY}^2(s,\nu),\\
F_{ZZ}^2(s) = {1 \over 2 M_s} \sum_{\nu=0}^{2 M_s - 1} f_{ZZ}^2(s,\nu),
\label{eq::variance.average}
\end{eqnarray}
in which all the boxes contribute to the sums with the same weights. The scale dependence of $\left[ F_{XY}^q(s) \right]^{1/q}$ and $\left[ F_{ZZ}^q(s) \right]^{1/q}$ can indicate a fractal character of the signals if it is power-law, but here we allow for any form of this dependence.

The $q$-dependent detrended cross-correlation coefficient $\rho_q(s)$ is defined by means of the $q$th order fluctuation functions~\cite{kwapien2015}:
\begin{equation}
\rho_q(s) = {F_{XY}^q(s) \over \sqrt{ F_{XX}^q(s) F_{YY}^q(s) }}.
\label{eq::rho.q}
\end{equation}
For $q=2$ Eq.~(\ref{eq::rho.q}) reduces to the definition of $\rho_{\rm DCCA}$~\cite{zebende2011}. The filtering ability of $\rho_q(s)$ manifests itself in such a way that the more deviated from the value $q=2$ the exponent $q$ is, the more extreme fluctuations in the corresponding boxes contribute to the coefficient $\rho_q(s)$. For $q \ge 0$, values of $\rho_q$ fit within the range
\begin{equation}
-1 \le \rho_q \le 1 .
\label{eq::normalized}
\end{equation}
As in the case of the Pearson and the $\rho_{\rm DCCA}$ coefficient, $\rho_q=1$ indicates a perfect correlation, $\rho_q=0$ indicates independent signals, and $\rho_q=-1$ indicates a perfect anticorrelation. However, for $q < 0$ the coefficient $\rho_q$ is not bound and for independent or weakly correlated signals it may happen that $|\rho_q|>1$. In such a case, an inverted value of $\rho_q$ is considered, which maps the coefficient back into the [-1,1] interval~\cite{kwapien2015}.

In a multivariate analysis of $N$ parallel signals, one has to deal with $N(N-1)/2$ different coefficients $\rho_q$ for each considered box size (time scale) $s$. It is thus convenient to put them in a matrix of size $N \times N$ that can be considered a matrix defining an $N$-node weighted network. Similar to the Pearson correlation coefficient, $\rho_q$ is not a metric, because it does not fulfill the triangle inequality. In the standard procedure of a minimum spanning tree construction, a matrix of the correlation coefficients is transformed into a (metric) distance matrix based on the formula: $d_{XY} = \sqrt{ 2(1-c_{XY}) }$, where $c_{XY}$ is the Pearson coefficient calculated for the signals $X$ and $Y$~\cite{mantegna1999}. However, an analogous transformation using $\rho_q$,
\begin{equation}
d_{XY}^{(q)} = \sqrt{ 2(1-\rho_q^{(XY)}) },
\label{eq::metric}
\end{equation}
might not produce a metric in general.

In order to test numerically if the triangle inequality: $d_{XY}^{(q)}(s)+d_{YZ}^{(q)}(s) \ge d_{XZ}^{(q)}(s)$ is valid for any triple of the signals $X,Y,Z$, we generate a set of $N=100$ signals of length $T=10^6$ being the autoregressive fractional moving average (ARFIMA) processes~\cite{hosking1981}. Next, for a given $s$ and $q$ we calculate $\rho_q$ for each pair of signals (4,950 pairs total) and transform its value into the distance $d_{XY}^{(q)}$. Then the inequality is tested for all possible distance triples (161,700 total) and the exceptions are counted. The same sequence of steps is repeated for different values of $q$ ($-10 \le q \le 10$). The results for sample values of $s$ and $q$ are collected in Tab.~\ref{tab::triangle.exceptions}. For $q \ge 1$, the distances always fulfill the inequality, which means that $d_{XY}^{(q)}$ behaves like a metric. However, for $q \le 0$ the exceptions do occur indicating that the distance cannot be considered a metric. We also applied the same test to randomized empirical data from a stock market (logarithmic price fluctuations of the 100 largest American companies, the same data set that will be studied in Sect.~\ref{sect.3}) and found qualitatively similar results: no violation of the triangle inequality for $q \ge 1$ and frequent violations for $q \le 0$. Therefore, our further analysis will be restricted to positive $q$'s. This, however, does not limit the robustness of our approach as, in many empirical systems, the small fluctuations that are selected by the negative values of $q$ are associated with noise.

\begin{table}
\begin{tabular}{|c||c|c|c|c|c|c|c|c|c|}
\hline
\backslashbox{$s$}{$q$} & \ -4.0 \hfill & \ -3.0 \hfill & -2.0 & -1.0 & \ \ 0.0 \hfill & \ \ 1.0 \hfill & \ \ 2.0 \hfill & \ \ 3.0 \hfill & \ \ 4.0 \hfill \\
\hline\hline
 20 & 0 & 3 & 2626 & 0 & 27 & 0 & 0 & 0 & 0 \\
\hline
 200 & 0 & 2 & 1984 & 80 & 0 & 0 & 0 & 0 & 0 \\
\hline
 2,000 & 0 & 49 & 2160 & 5162 & 0 & 0 & 0 & 0 & 0 \\
\hline
 20,000 & 52 & 252 & 1955 & 9089 & 0 & 0 & 0 & 0 & 0 \\
\hline
\end{tabular}
\caption{Number of exceptions from the triangle inequality for 161,700 triples of the $d_{XY}^{(q)}(s)$ distances calculated for the 100 uncorrelated time series of the ARFIMA process.}
\label{tab::triangle.exceptions}
\end{table}

Now we are prepared to define a $q$-dependent minimum spanning tree by using the $q$-distance given by Eq.~(\ref{eq::metric}). $q$MST can be constructed by applying the Kruskal's algorithm to $d_{XY}^{(q)}$. This algorithm consists of two essential steps: (1) sorting the elements of a distance matrix from the smallest to the largest, and (2) going from the smallest distances and connecting not yet connected nodes to the closest ones. After connecting all the nodes, the resulting tree consists of $N$ nodes and $N-1$ weighted edges. It is so optimized that the sum of all the distances $d_{XY}^{(q)}$ is minimum possible. The $q$MSTs inherit the properties of $\rho_q$, so by choosing a particular value of $q$, the related tree expresses only those correlations that are present among the fluctuations of a selected size: the large ones for $q \gg 2$ and the small ones for $q < 2$. However, even though for $q=2$ this procedure of constructing the detrended minimal spanning trees gives the same weights to all fluctuation amplitudes, one has to remember that even for this value of $q$ it substantially differs from the standard procedure that does not involve detrending.

\section{$q$MST${\rm s}$ for empirical data}
\label{sect.3}

In order to present an example of a $q$MST analysis, we consider a set of time series representing logarithmic stock-price fluctuations (returns):
\begin{equation}
r_X(t,\Delta t))=\ln p_X(t+\Delta t) - \ln p_X(t)
\end{equation}
for the $N=100$ largest American companies traded on the New York Stock Exchange over the years 1998-1999~\cite{tickers}. We choose sampling interval of $\Delta t=1$ min. to obtain sufficiently long time series ($T=203,190$). We calculate $\rho_q(s)$ for all possible pairs of stocks, a few different time scales $s$: 20 min., 60 min., 390 min. (a trading day), 1950 min. (a trading week), and 7800 min. (approx. a trading month), and a few different values of $q$ from $q=1.0$ to $q=6.0$ (although we use only integer values, arbitrary positive real values may also be used). Next, we derive the metric distances $d_{XY}^{(q)}(s)$ and applied the Kruskal's algorithm to obtain a minimum spanning tree for each value of $s$ and $q$ (36 trees total).

\begin{figure}[t]
\includegraphics[scale=0.15]{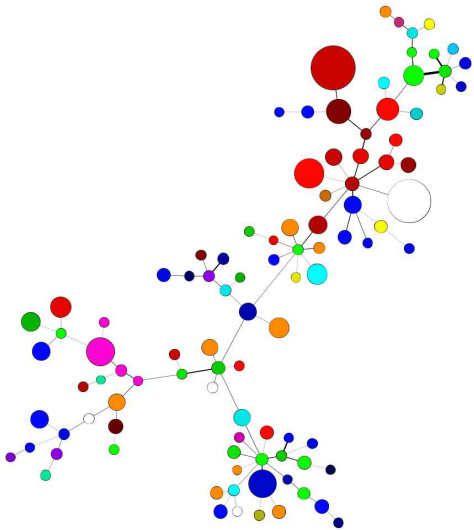}
\caption{(Color online) The $q$-dependent minimum spanning tree example ($s=20, q=5.0$). Symbol size is proportional to the market capitalization of a stock on Dec 31, 1999, while different sectors are denoted by different colors and different subsectors within the same sector are denoted by shades of a particular color: technology (from light red to dark brown), services (from light blue to dark blue), energy (from light magenta to dark magenta), financial (from light green to dark green and olive), consumer non-cyclical (from cyan to turquoise), basic materials (from yellow to greenish yellow), healthcare (orange and light brown), consumer cyclical (violet), capital goods (seagreen), utilities (violetred), and conglomerates (white). Edge thickness is proportional to $\rho_q$ (the thicker it is, the more correlated the stocks are).}
\label{fig::s20.complete}
\end{figure}

Fig.~\ref{fig::s20.complete} displays the $q$MST for the shortest time scale $s=20$ and for sample exponent $q=5.0$. Different industrial sectors and subsectors are distinguished by different colors and color shades, respectively, and the node symbols' size is proportional to the market capitalizaton of the corresponding company. The higher is the calculated $\rho_q^{(XY)}$, the thicker is an edge connecting the nodes $X$ and $Y$. One can see that the tree topology is somewhere between a centralized and a distributed network. There are hubs with a degree up to $k=9$ and a number of peripheral nodes with $k=1$, but the shortest path length is rather high: $L\approx 7.9$. The edge weights are also rather small with a number of edges denoted by the lines as thin as possible. Several industrial sectors are visible as clusters, but there is a number of nodes that are connected to nodes from different sectors. This suggests that there is a significant randomness in the distribution of the nodes across this tree. An immediate question follows whether these effects are genuine or they are rather statistically insignificant and pertinent to noise.

We therefore have to estimate the statistical significance of the edge weights, at least approximately. Since our data is strongly leptokurtic~\cite{plerou1999,drozdz2003,drozdz2007,rak2013}, we define the null hypothesis stating that the $\rho_q(s)$ values can fully be explained by the cross-correlations between random processes with the same p.d.f. as the empirical data under study. To test the results against this hypothesis, we shuffle all the time series 50 times and create 50 independent sets of surrogate data. Then for each set we derive all possible values of $\rho_q^{(XY)}(s)$ for the same choices of $s$ and $q$ as we did before and identify the maximum values of $\rho_q$ in each case. Next, for a given $s$ and $q$, we collect all the maximum coefficients and calculate their mean $\bar{\rho_q}$ and standard deviation $\sigma_{\rho}$. Finally, we assume gaussianity of the p.d.f. and set a threshold on $\tau_{\rho} = \bar{\rho_q} + 2 \sigma_{\rho}$. If for some $X,Y$ we obtain $\rho_q^{(XY)}(s) \ge \tau_{\rho}$, we consider the null hypothesis to be rejected in this case and not rejected otherwise. The specific thresholds for different values of $s$ and $q$ are gathered in Tab.~\ref{tab::thresholds}.

\begin{table}
\begin{tabular}{|c||c|c|c|c|c|c|}
\hline
\backslashbox{$s$}{$q$} & \ \ 1.0 \hfill & \ \ 2.0 \hfill & \ \ 3.0 \hfill & \ \ 4.0 \hfill & \ \ 5.0 \hfill & \ \ 6.0 \hfill \\
\hline\hline
 20 & \ 0.0155 \ & \ 0.0114 \ & \ 0.0179 \ & \ 0.0235 \ & \ 0.0248 \ & \ 0.0255 \ \\
\hline
 60 & \ 0.0261 \ & \ 0.0184 \ & \ 0.0223 \ & \ 0.0296 \ & \ 0.0316 \ & \ 0.0309 \ \\
\hline
 390 & \ 0.0733 \ & \ 0.0507 \ & \ 0.0486 \ & \ 0.0607 \ & \ 0.0703 \ & \ 0.0740 \ \\
\hline
 1950 & \ 0.1493 \ & \ 0.1032 \ & \ 0.0841 \ & \ 0.0837 \ & \ 0.0905 \ & \ 0.0971 \ \\
\hline
 7800 & \ 0.3129 \ & \ 0.2183 \ & \ 0.1840 \ & \ 0.1842 \ & \ 0.1948 \ & \ 0.2029 \ \\
\hline
\end{tabular}
\caption{Thresholds $\tau_{\rho}=\bar{\rho_q}+2*\sigma_{\rho_q}$ representing the $\rho_q(s)$ values above which we reject the null hypothesis of random correlations.}
\label{tab::thresholds}
\end{table}

\begin{figure}[t]
\begin{center}
\includegraphics[scale=0.08]{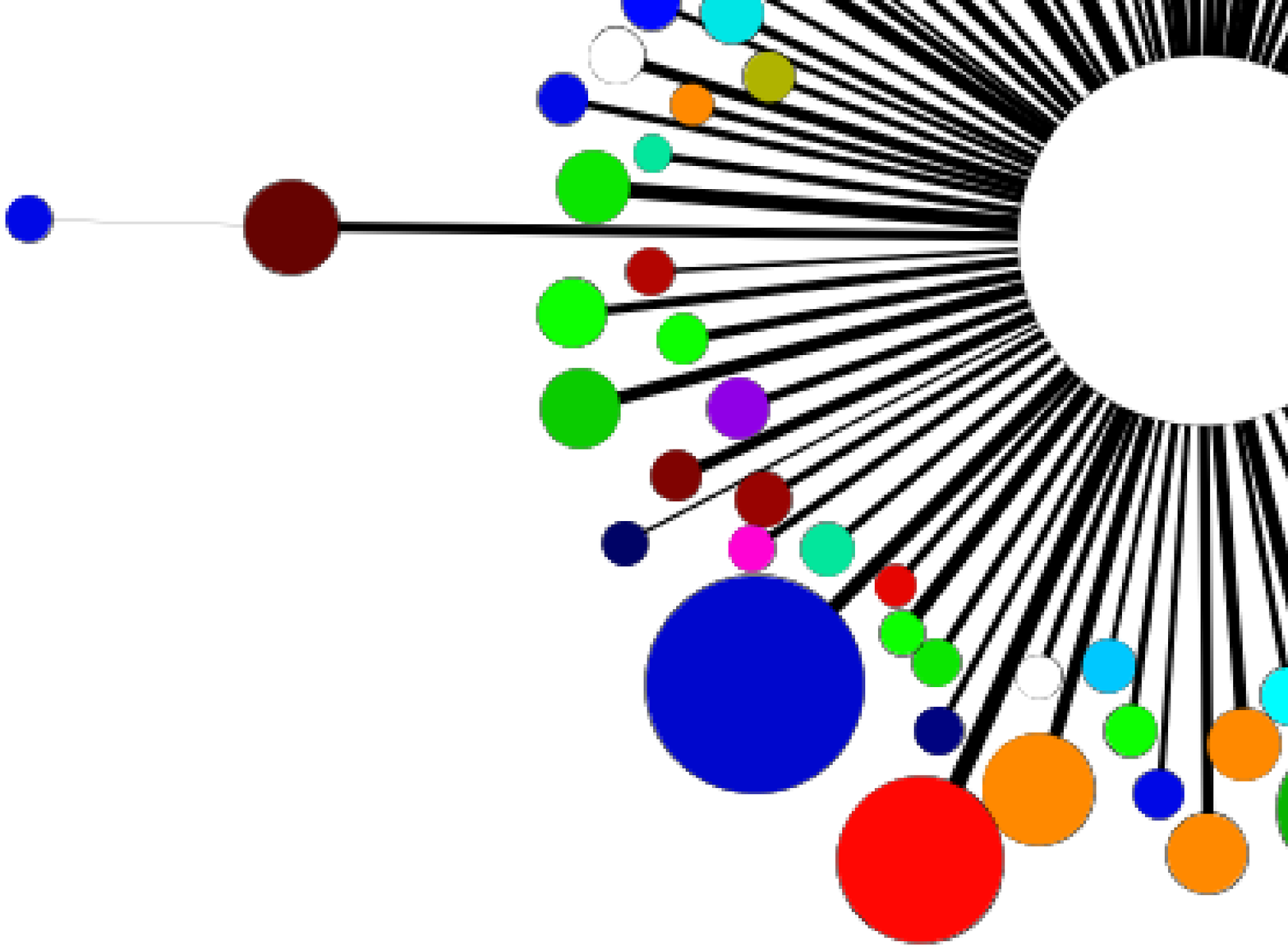}
\hspace{0.4cm}
\includegraphics[scale=0.08]{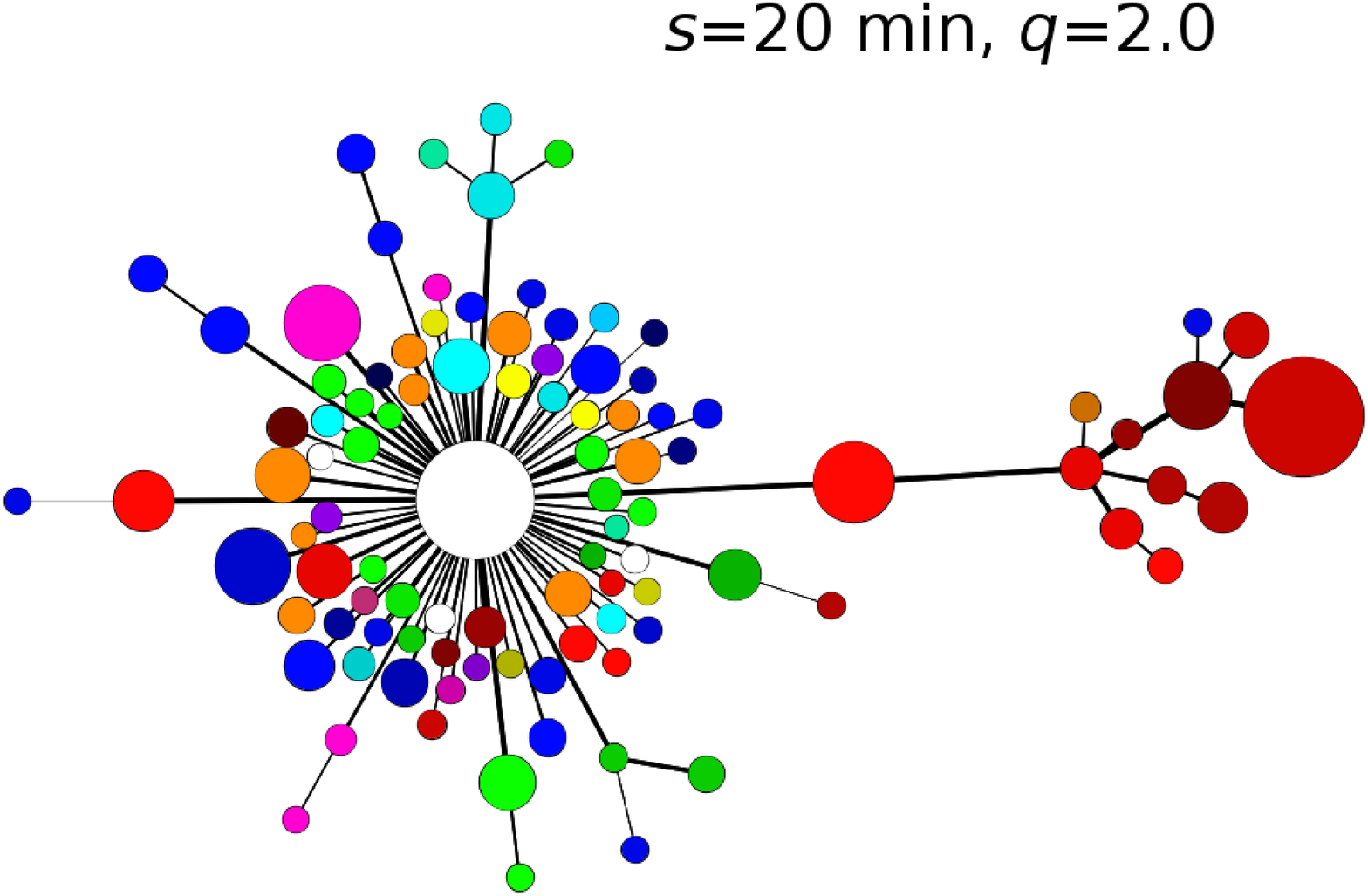}

\includegraphics[scale=0.08]{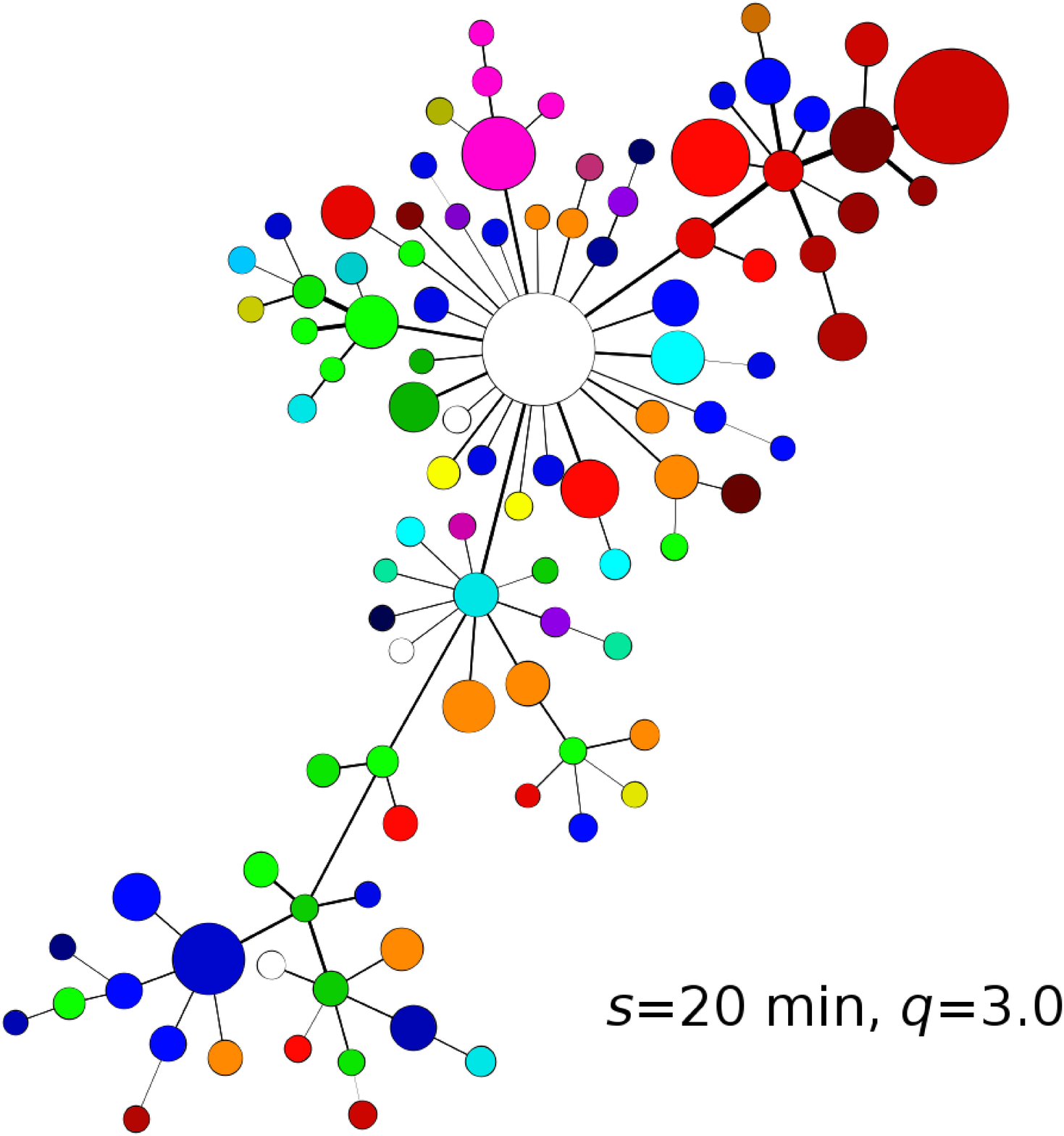}
\includegraphics[scale=0.08]{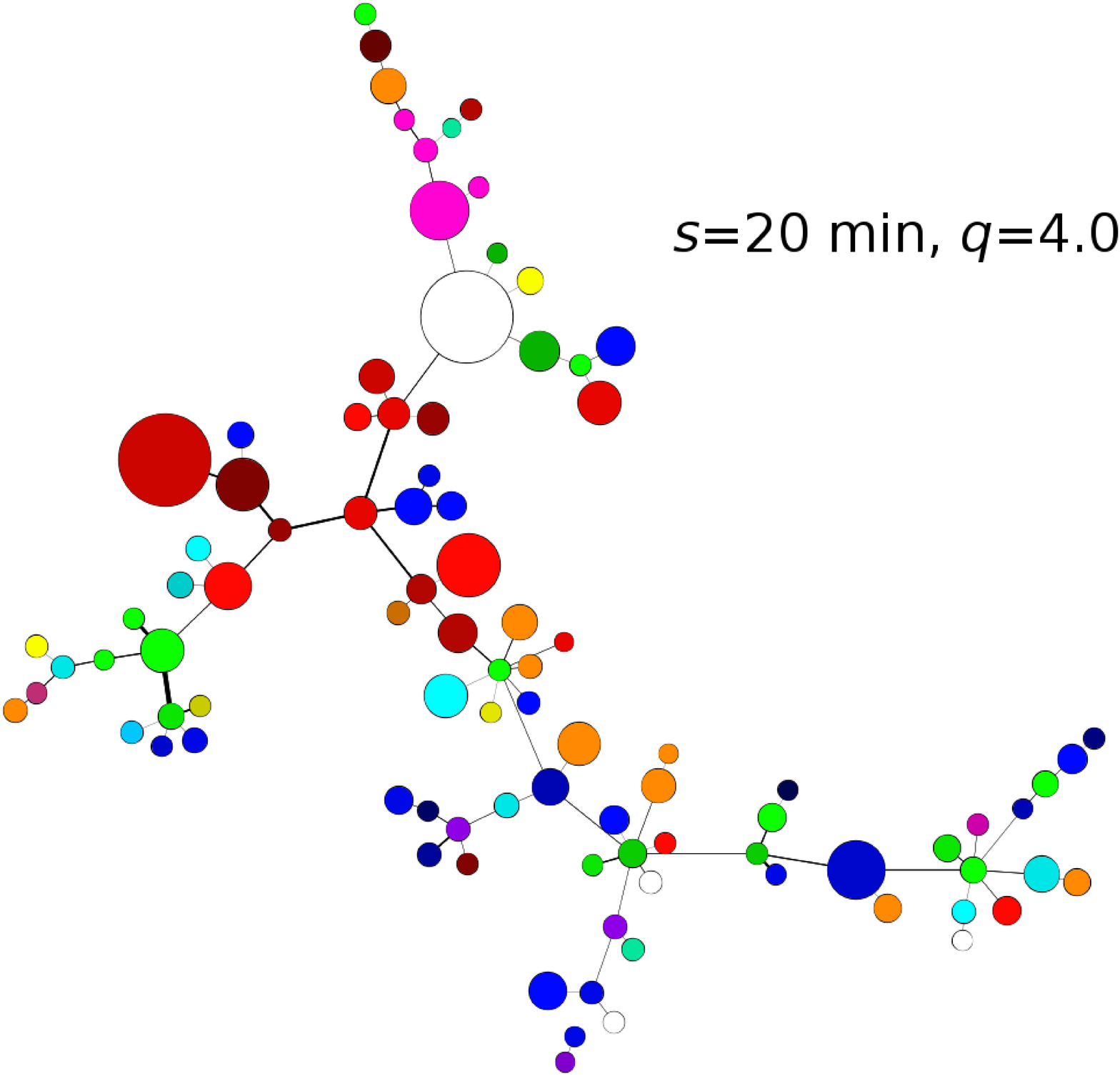}

\includegraphics[scale=0.08]{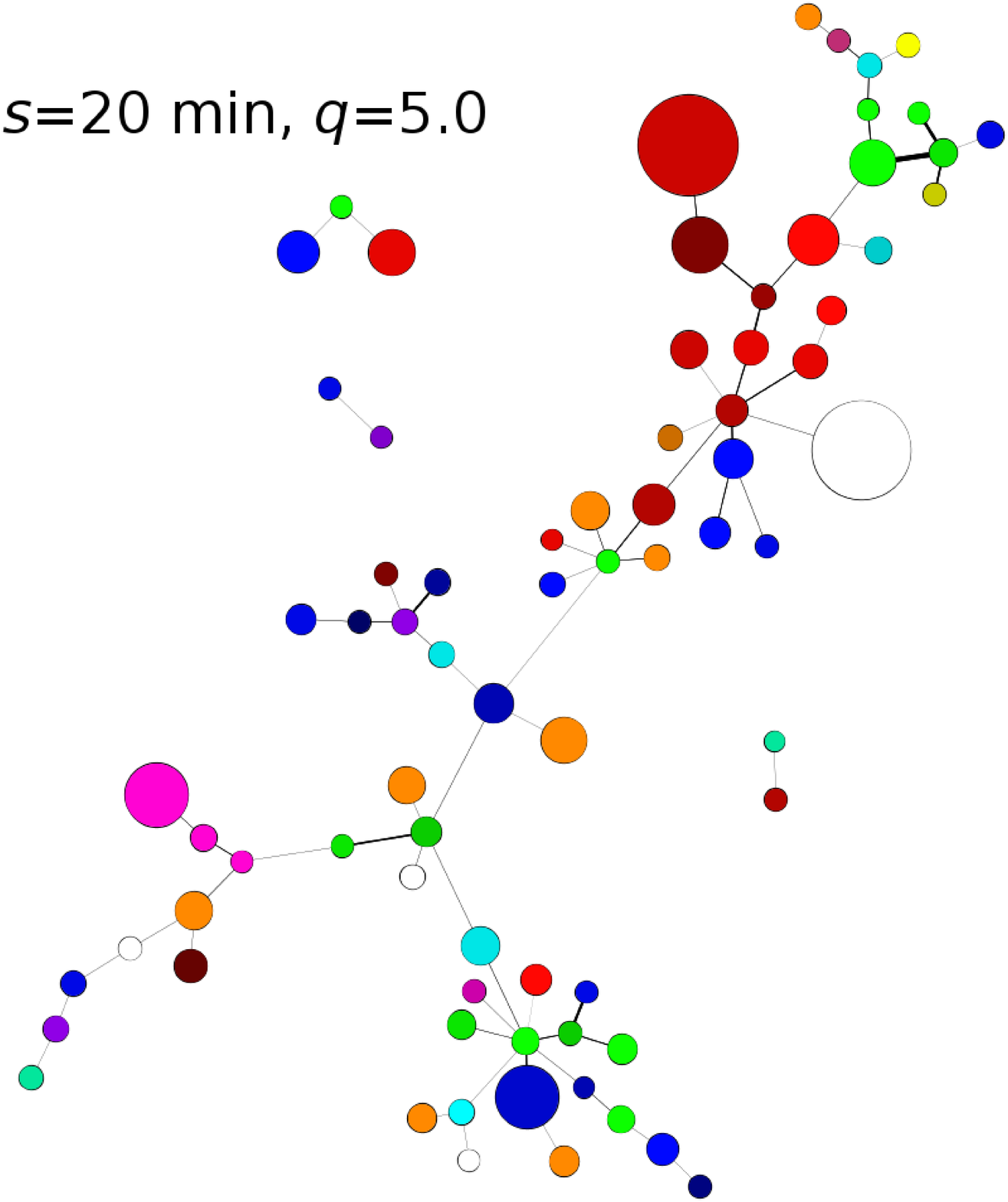}
\includegraphics[scale=0.08]{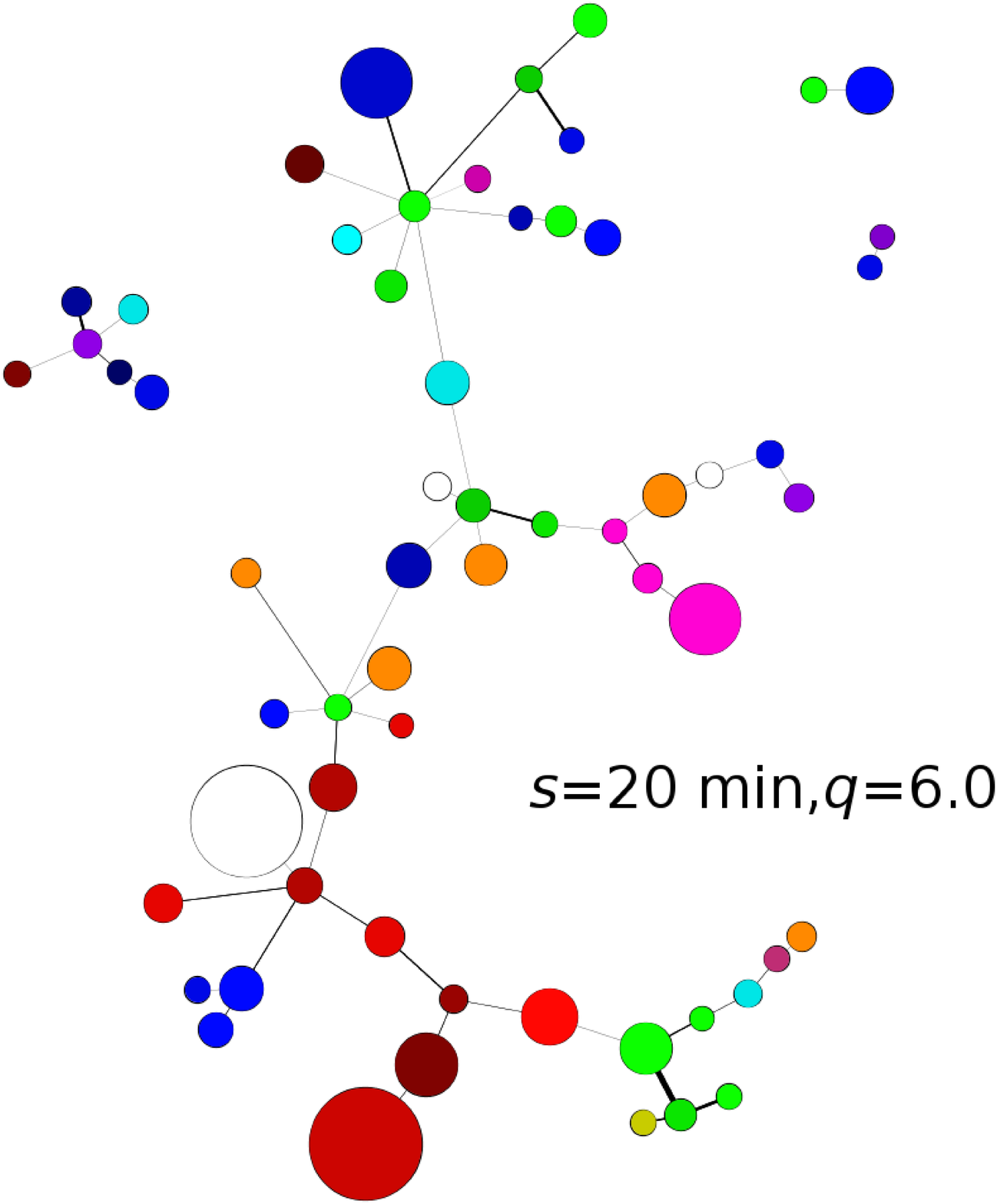}
\end{center}
\caption{(Color online) The $q$-dependent minimum spanning trees formed for $s=20$ and for a few different values of $q$ from 1.0 to 6.0 (from left to right and from top to bottom). The trees have been filtered to remove statistically insignificant edges ($\rho_q < \tau_{\rho}$) and nodes with degree $k=0$. (Differences in size of the node symbols between the trees are artifact.)}
\label{fig::s20.q1-6}
\end{figure}

The so-filtered $q$MST graphs for $s=20$ are shown in Fig.~\ref{fig::s20.q1-6} ordered by growing $q$. Three observations can be drawn from these pictures. First, the tree transforms itself from a highly centralized structure ($L \approx 2.2$) with a dominant hub (General Electric) of degree $k=87$ and a secondary hub (Cisco) with $k=10$ for $q=1.0$ to a distributed, random-like structure ($L \approx 8.2$) with a few regional hubs with $k > 4$: Yahoo! ($k=5$), Bank of New York ($k=6$), Morgan Stanley ($k=5$), and Chase Manhattan ($k=8$) for $q=6.0$. This topologic transition occurs gradually with increasing $q$. Filtering out the insignificant edges has no effect for $q=1.0$ as the tree consists of all the 100 nodes, but the effect is striking for $q=6.0$ where the structure is disconnected with only 50 nodes forming the main component (the main tree), while the remaining nodes either form three smaller components: one with $n=6$ nodes and two with $n=2$ nodes or do not form any connections (40 nodes). From the market perspective, a centralized structure with large edge weights corresponds to a global coupling among the stocks that causes the stock prices to move collectively. Their evolution can thus be described by a one-factor model. In a correlation matrix analysis this factor manifests itself as the distant largest eigenvalue and it is often called the market factor. On the other hand, a distributed topology with a few hubs of a small degree and many disconnected nodes indicates that the market can be decomposed into clusters often related to the industrial sectors and that many stocks have independent dynamics. The eigenvalue spectrum of a correlation matrix would show a few non-random eigenvalues and a bulk of random ones in this case~\cite{drozdz2001,plerou2002,kwapien2002,kwapien2012,wang2013b}.

The second observation is that, for the 20-min. time scale, the periods (the MFCCA boxes) characterized by the fluctuations of moderate amplitude are universally correlated among the stocks, while the periods with the largest fluctuations are correlated only within small groups of stocks and typically they are uncorrelated. This can be understood if one realizes that it needs time to develop large movements of the  whole market and 20 min. is too short to accomplish this.

Finally, the third observation from Fig.~\ref{fig::s20.q1-6} is that some node clusters are notably overlapping with the industrial sectors, especially for $3.0 \le q \le 5.0$. It is the most evident for the IT cluster (red, this cluster is seen even for $q=1.0$), the financial cluster (green), and the basic materials cluster (magenta). For the remaining sectors, their stocks are distributed across the $q$MST for $q \ge 3.0$. It is also noteworthy that since the edges shown in the filtered $q$MSTs are statistically significant, any restructuring of the connections while changing $q$ can be related to uncovering some previously ``hidden'' correlations.

\begin{figure}[t]
\begin{center}
\includegraphics[scale=0.08]{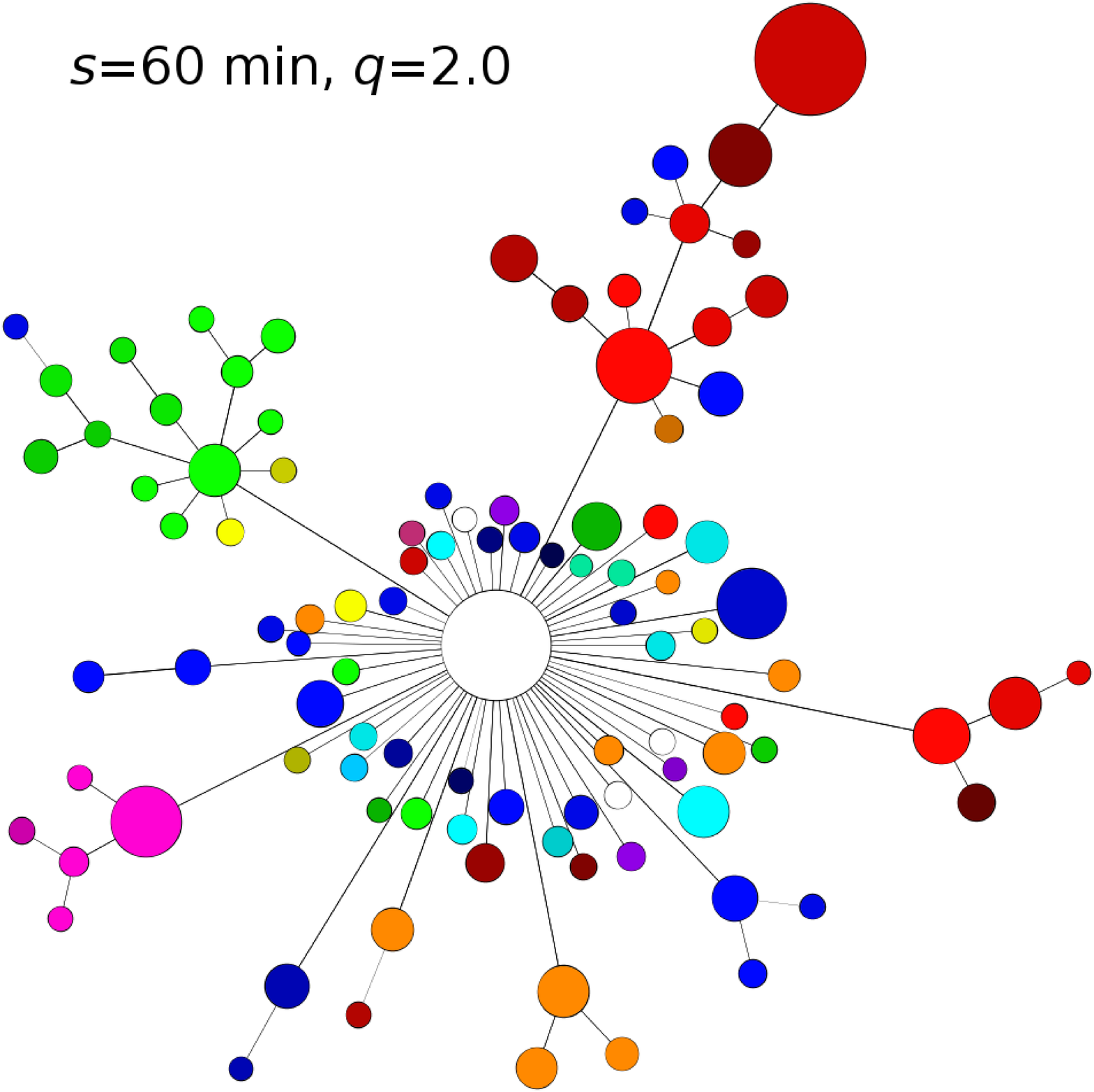}
\includegraphics[scale=0.08]{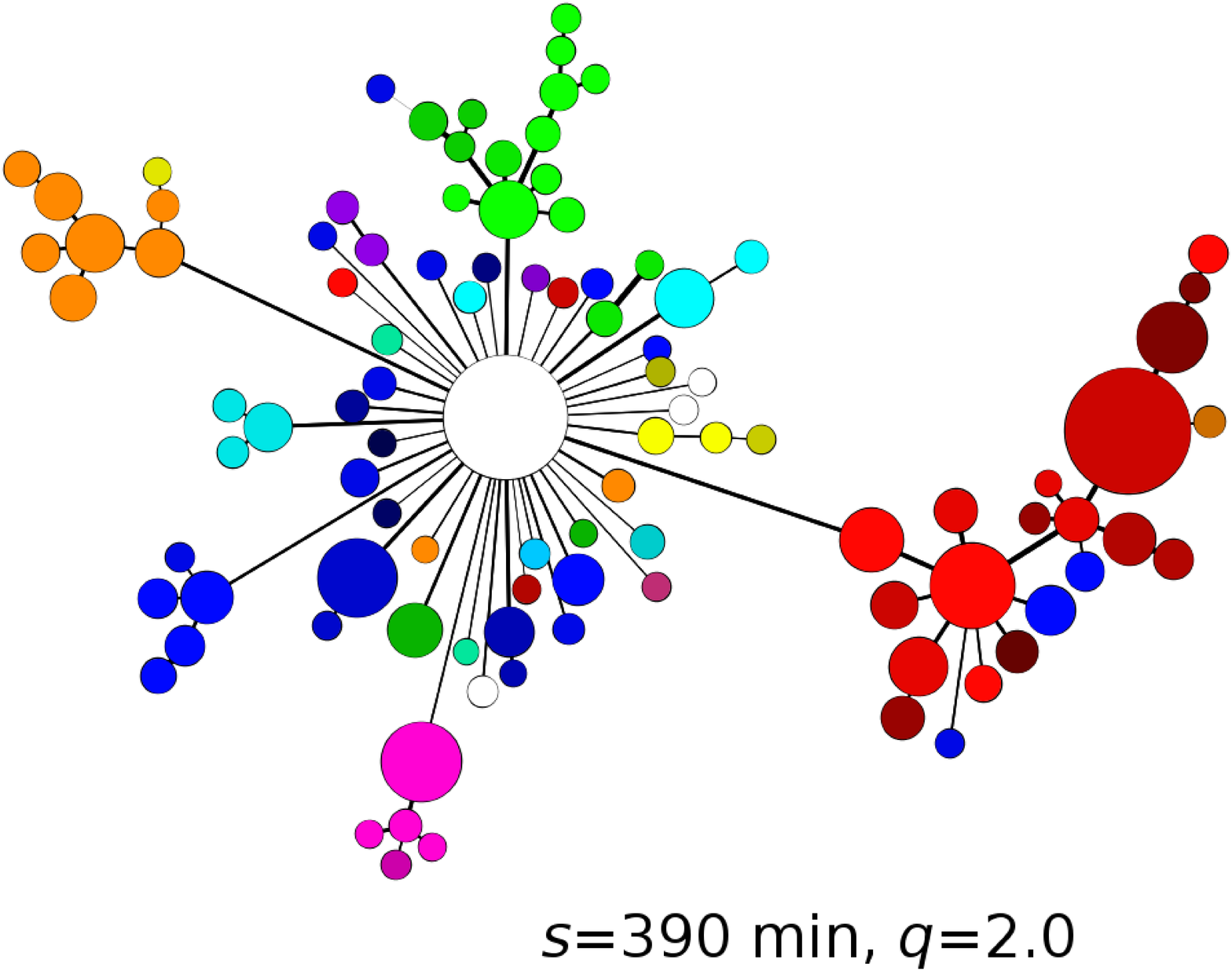}

\includegraphics[scale=0.08]{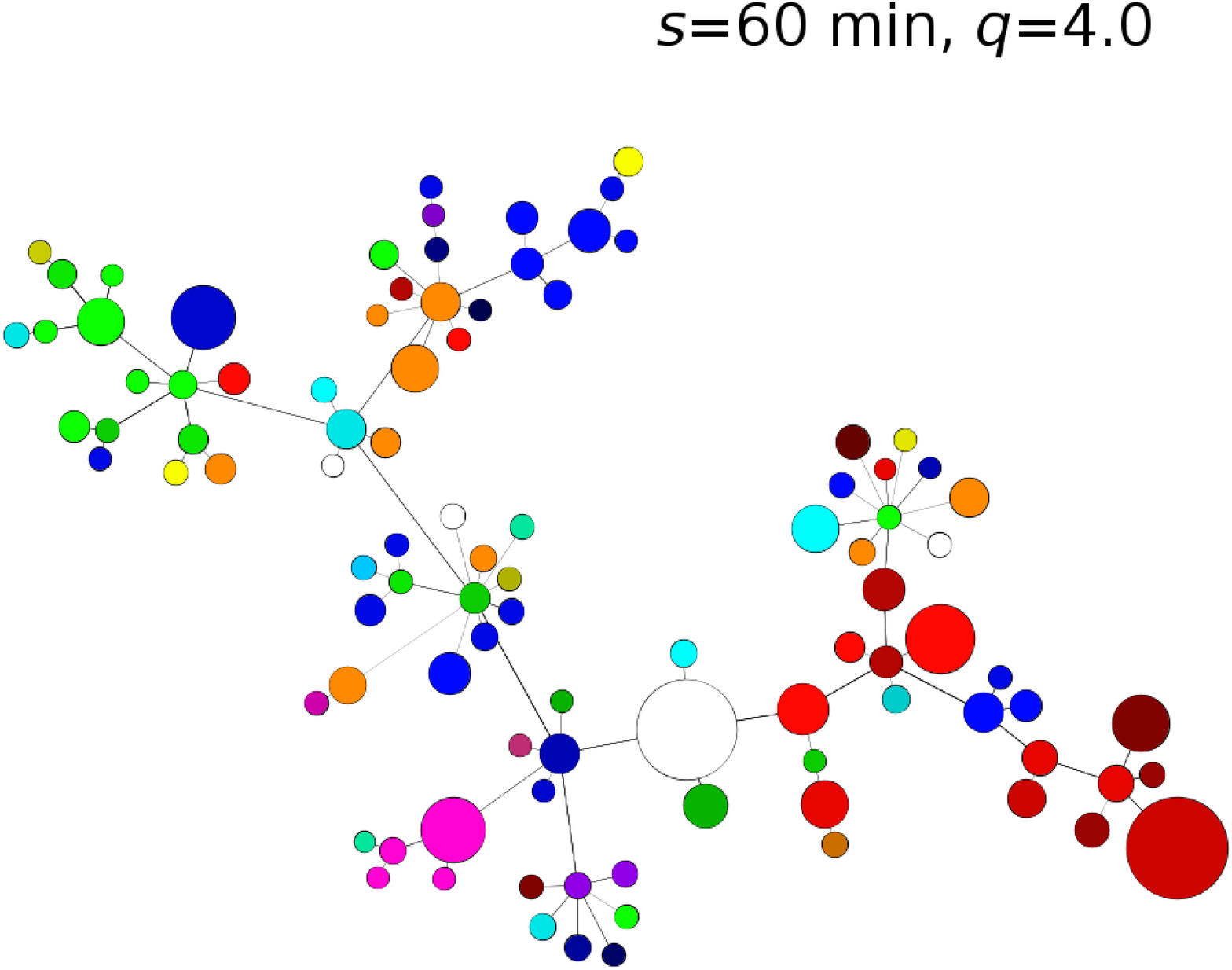}
\includegraphics[scale=0.08]{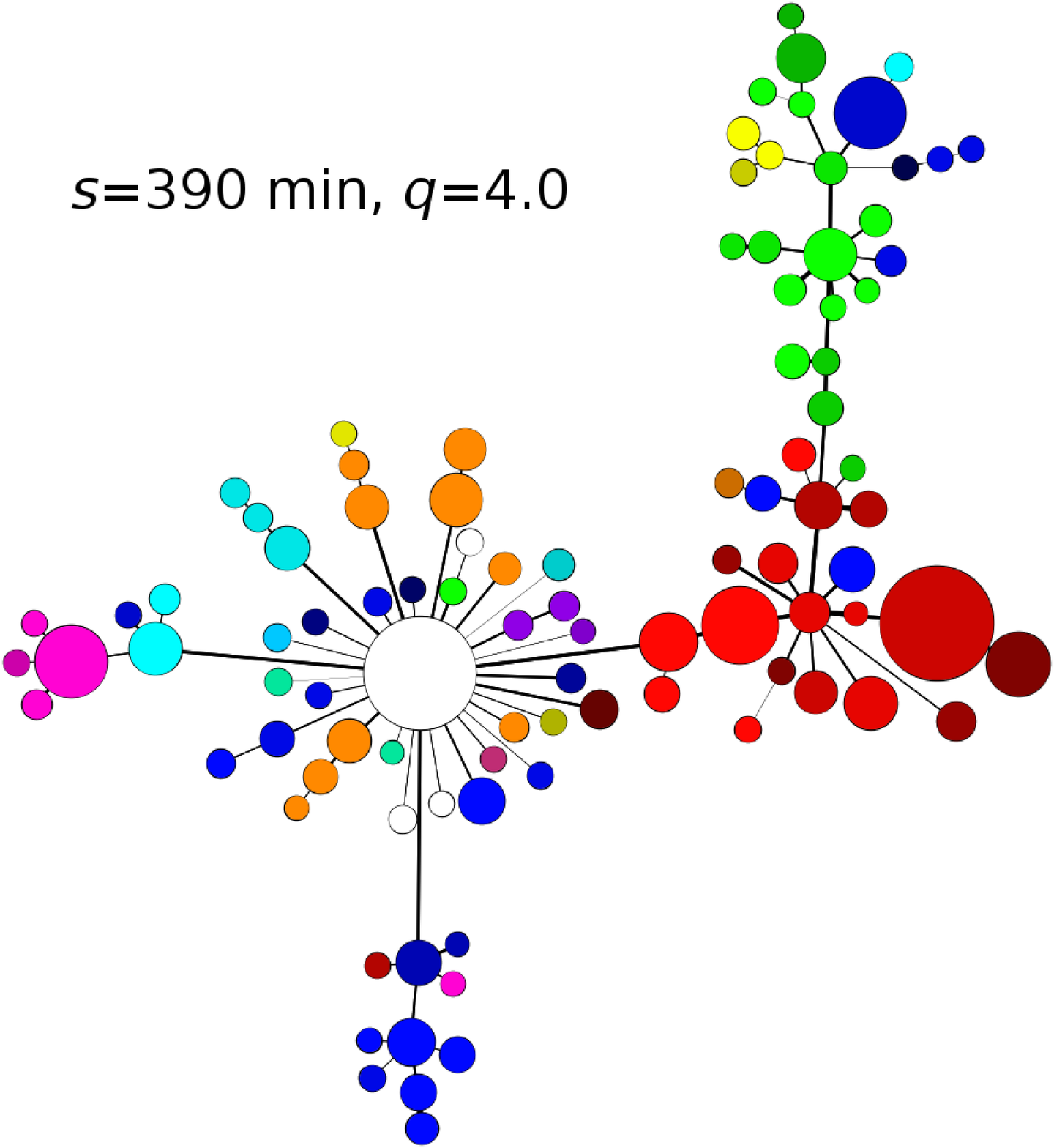}

\includegraphics[scale=0.08]{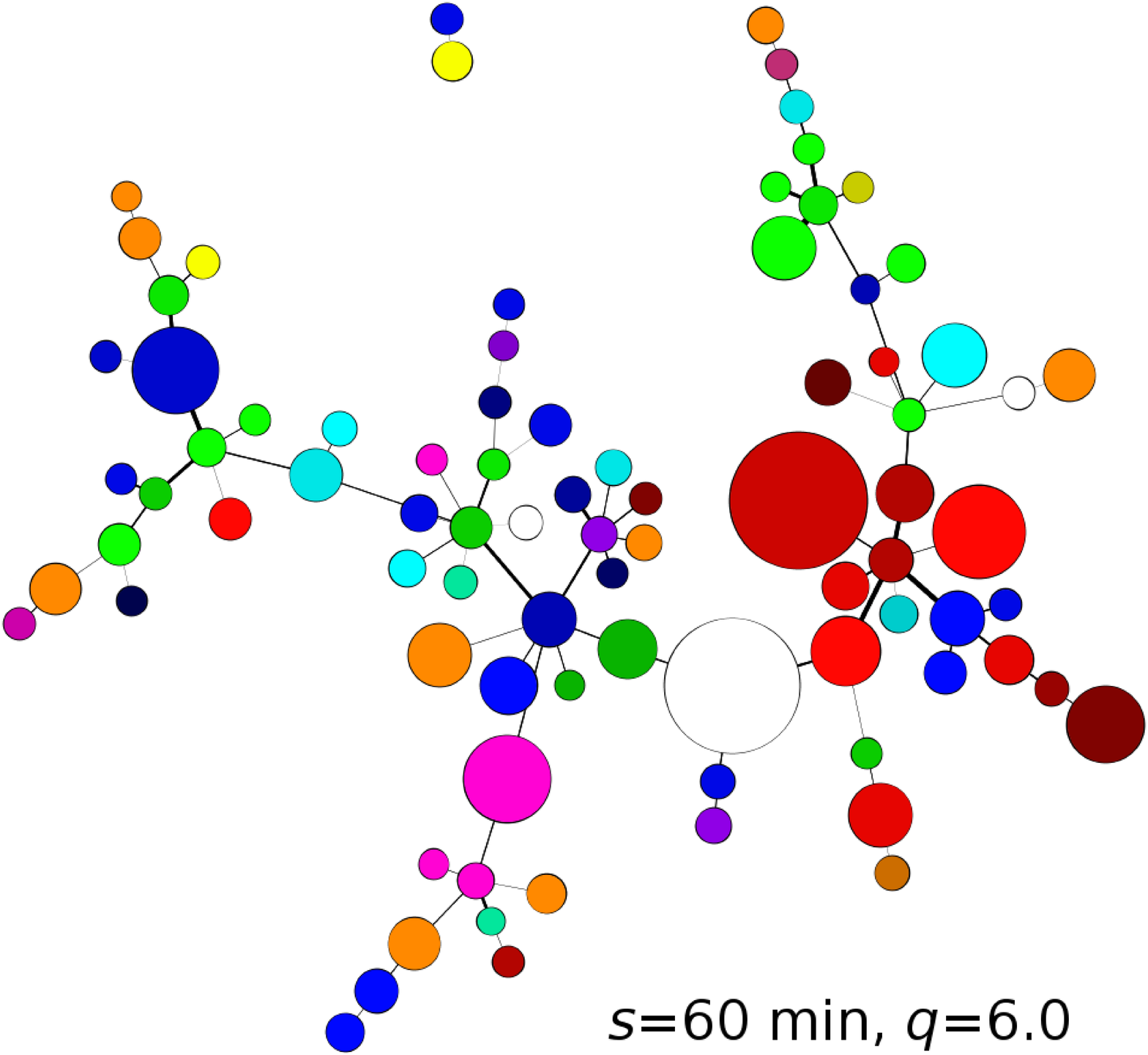}
\includegraphics[scale=0.08]{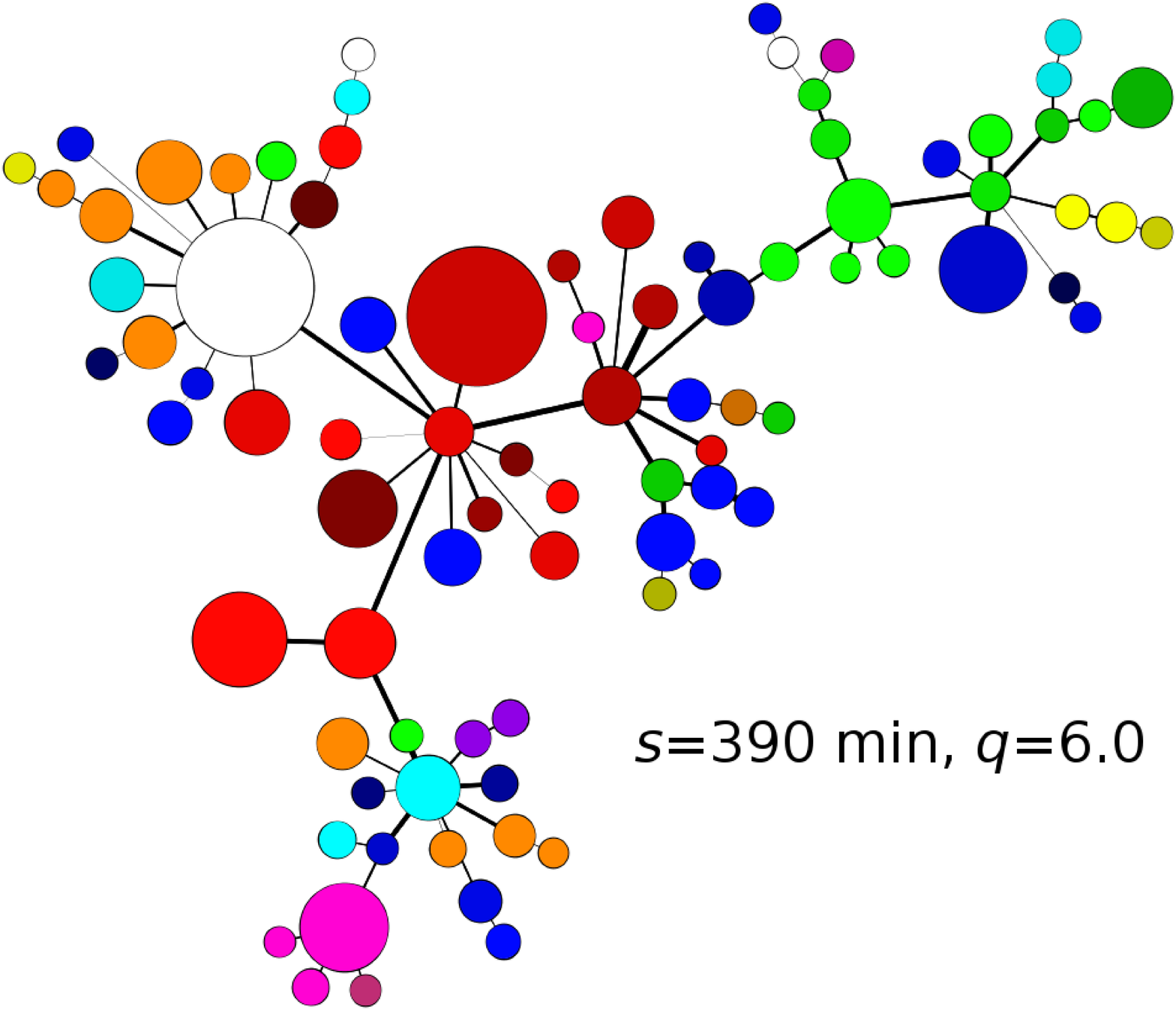}
\end{center}
\caption{(Color online) The $q$-dependent minimum spanning trees created for the hourly ($s=60$ min., left) and daily ($s=390$ min., right) time scales and for $q=2.0$ (top), $q=4.0$ (middle), and $q=6.0$ (bottom). The trees were filtered to remove statistically insignificant edges ($\rho_q < \tau_{\rho}$) and nodes with degree $k=0$.}
\label{fig::s60.s390.q2-6}
\end{figure}

\begin{figure}[t]
\begin{center}
\includegraphics[scale=0.08]{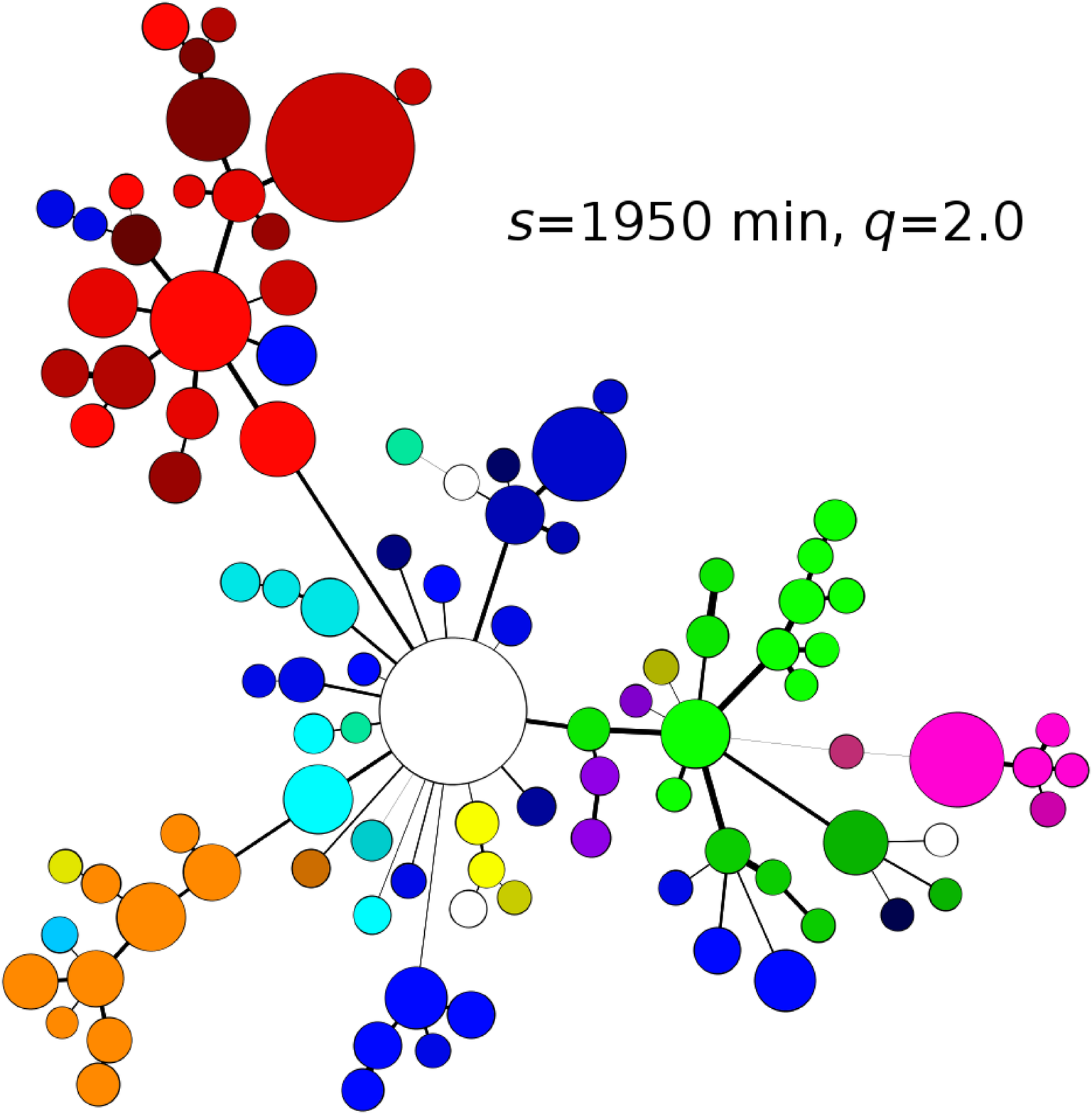}
\includegraphics[scale=0.08]{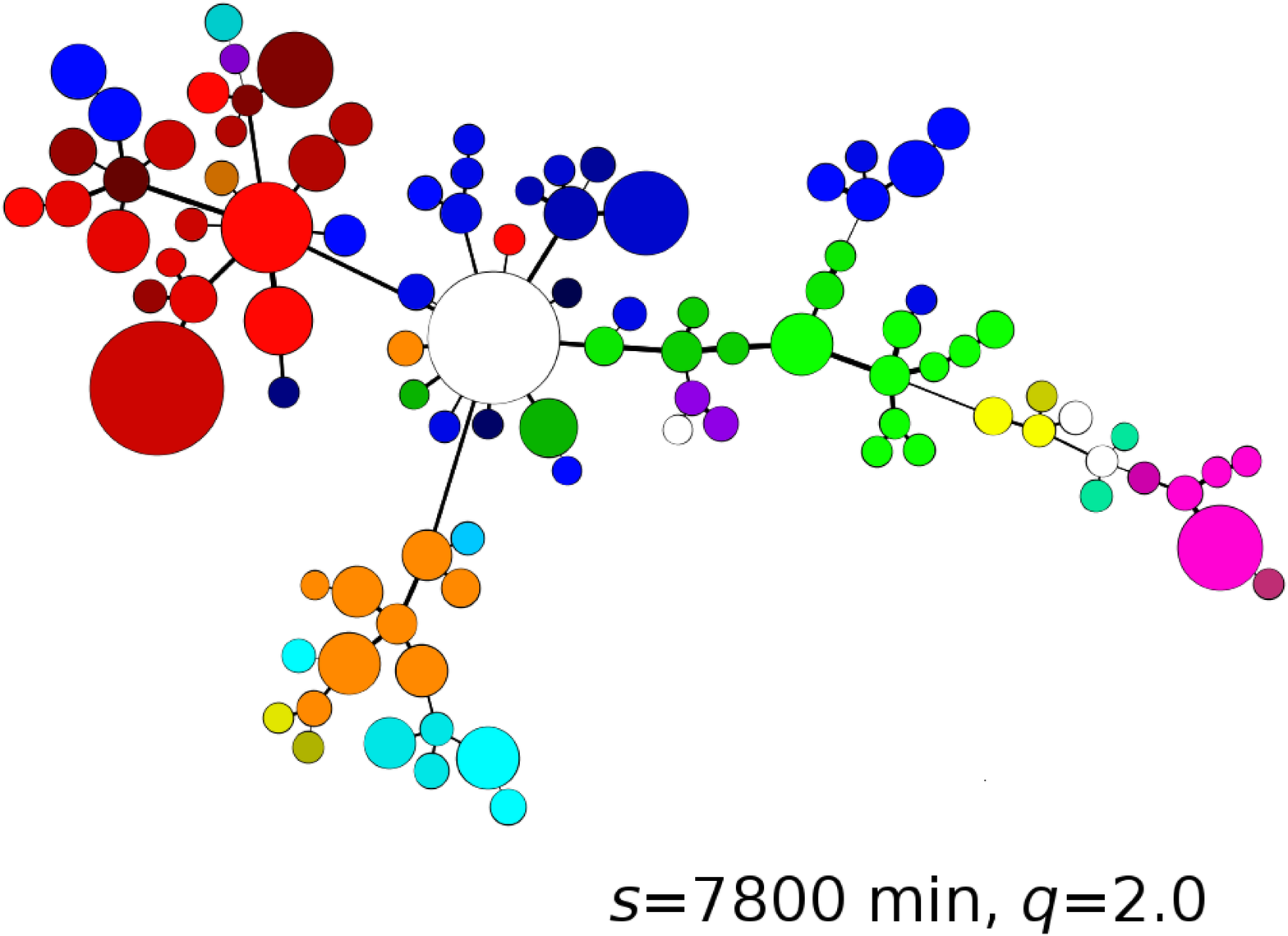}

\includegraphics[scale=0.08]{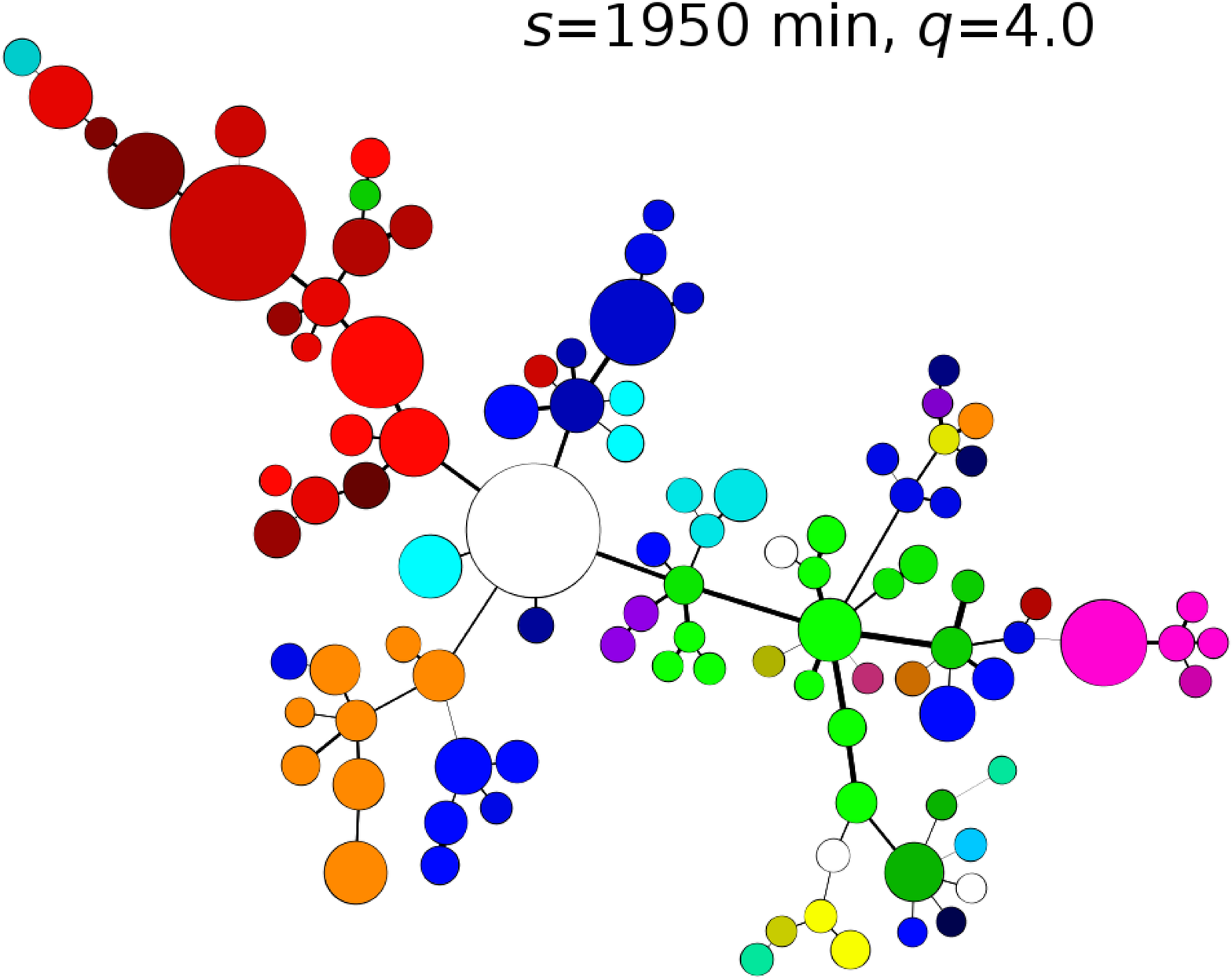}
\includegraphics[scale=0.08]{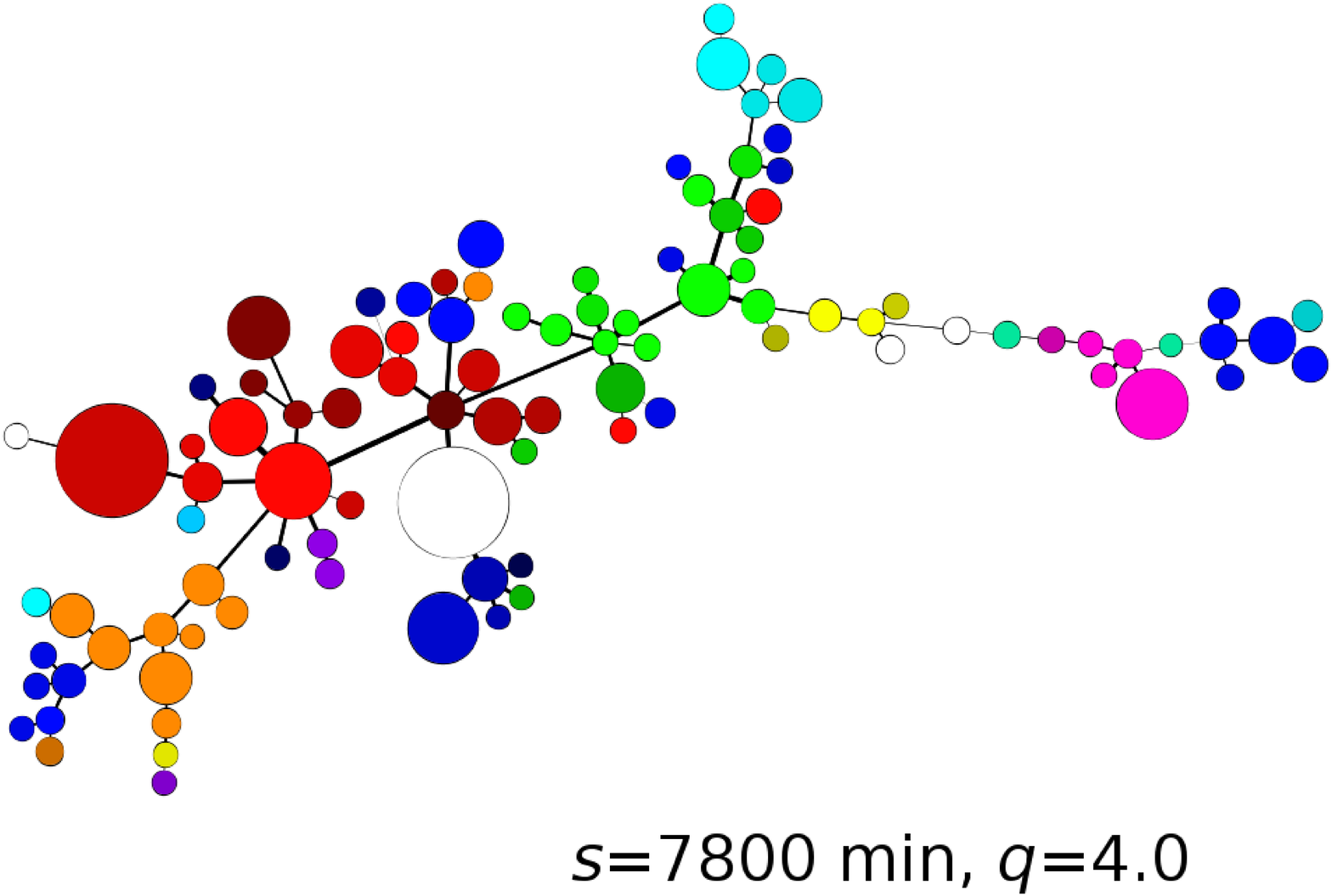}

\includegraphics[scale=0.08]{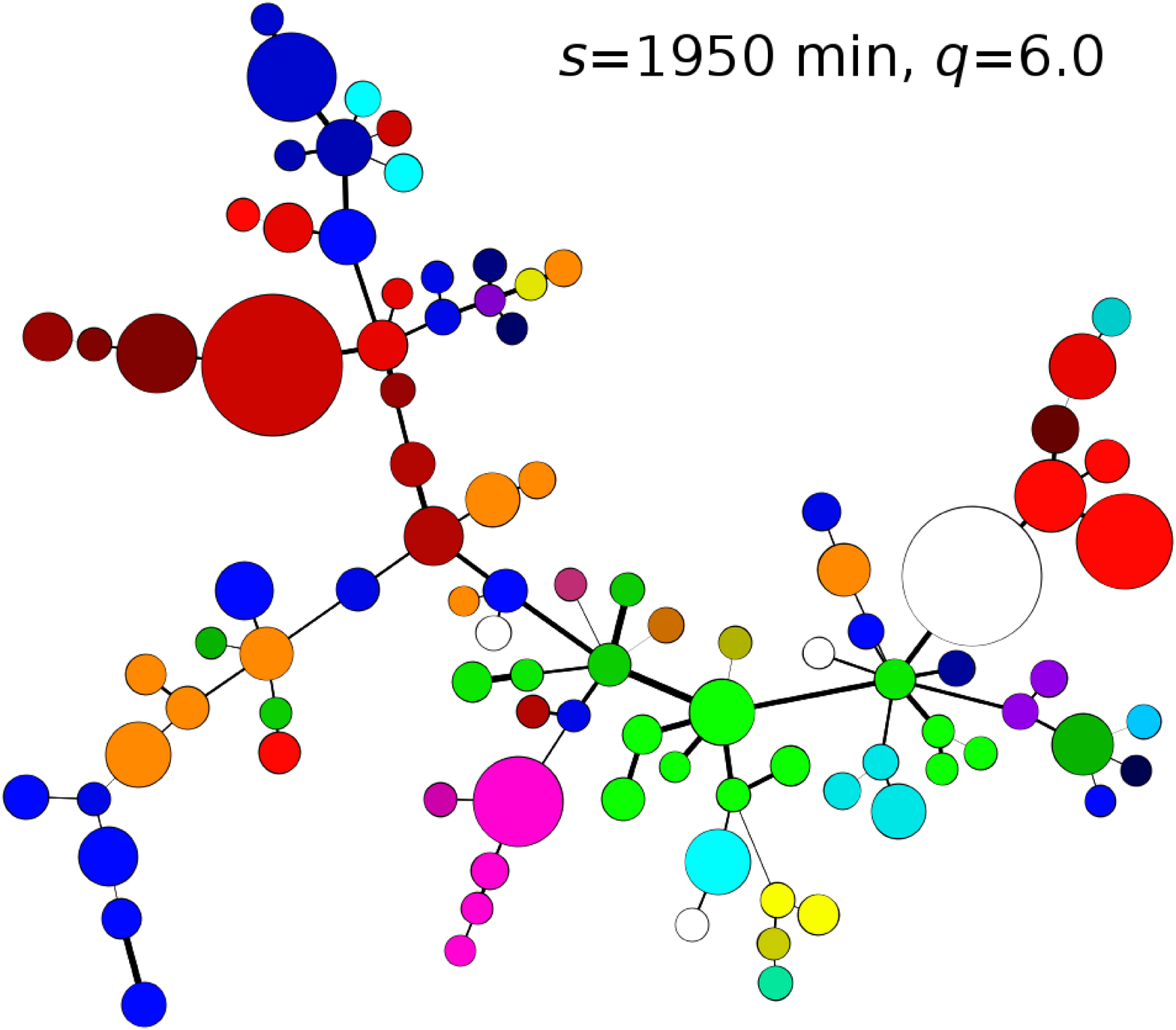}
\includegraphics[scale=0.08]{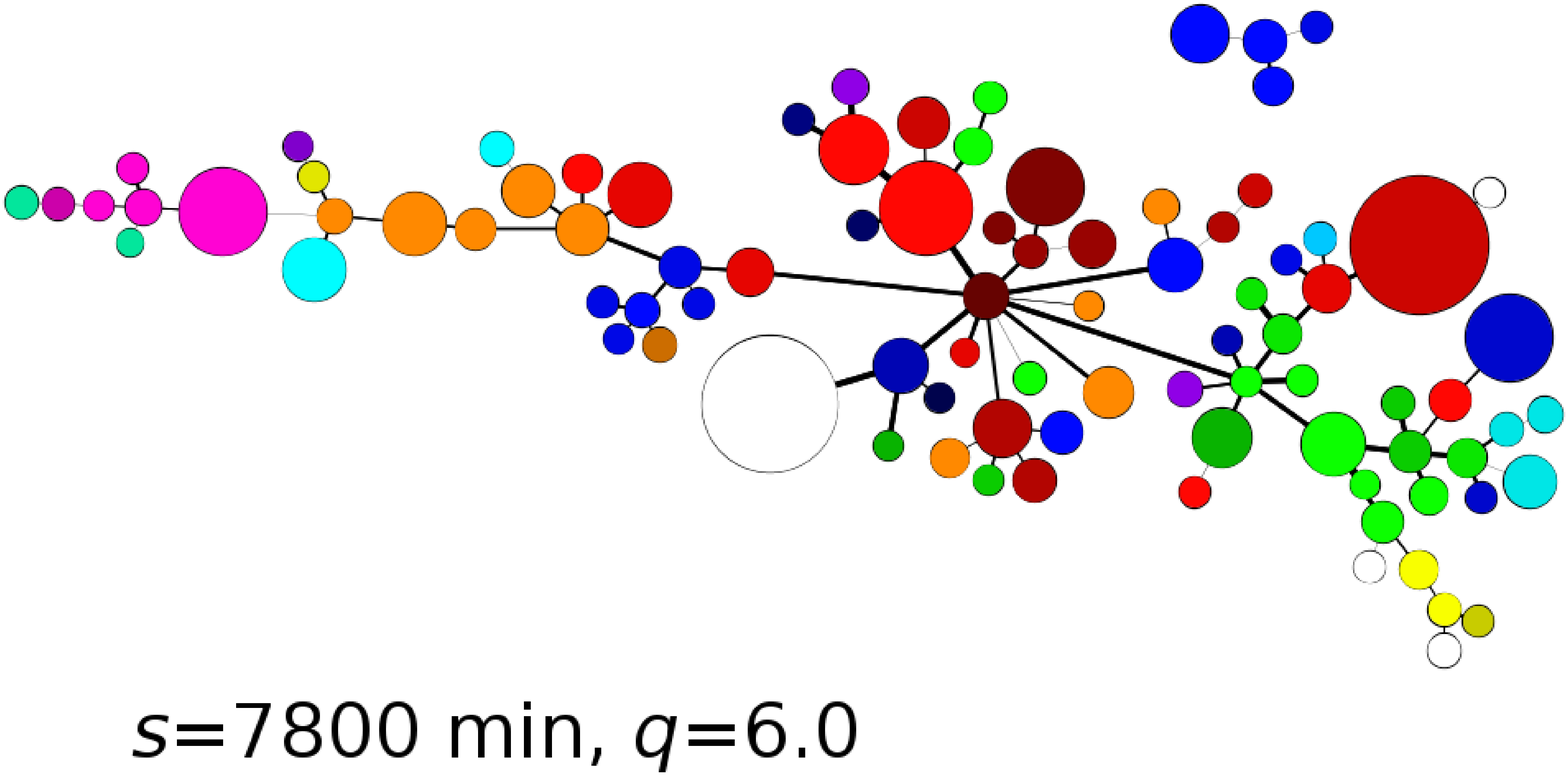}
\end{center}
\caption{(Color online) The $q$-dependent minimum spanning trees created for the weekly ($s=1950$ min., left) and monthly ($s=7800$ min., right) time scales and for $q=2.0$ (top), $q=4.0$ (middle), and $q=6.0$ (bottom). The trees were filtered to remove statistically insignificant edges ($\rho_q < \tau_{\rho}$) and nodes with degree $k=0$.}
\label{fig::s1950.s7800.q2-6}
\end{figure}

These results can be compared with those obtained for other time scales: $s=60$ min. and $s=390$ min. (Fig.~\ref{fig::s60.s390.q2-6}) as well as for $s=1950$ min. and $s=7800$ min. (Fig.~\ref{fig::s1950.s7800.q2-6}). For any given $s$, if we augment $q$, the main hub (being the General Electric stock) seen for $q=2.0$ loses its centrality and typically becomes a peripheral one. The other, less important hubs can either preserve their degree or lose it as well, which depends on $s$ or individually on a node. An additional general rule is that the shorter the time scale is, the more dramatic changes in topology can be observed while increasing $q$. The edge weights gradually become higher while increasing $s$ (the edges become thicker) for all the exponents $q$, which means that the average correlation strength among the stock pairs also increases. This remains in a perfect agreement with the results obtained earlier with other means suggesting existence of the so-called Epps effect~\cite{epps1979,kwapien2004,toth2005}.

It is interesting to observe changes in the tree topology for a fixed $q$ and variable $s$. For example, for $q=2.0$ there is a clear, gradual transition from the centralized one-factor structure for $s=20$ min. (top right in Fig.~\ref{fig::s20.q1-6}) to an almost perfect sectorial structure for the monthly scale (top right in Fig.~\ref{fig::s1950.s7800.q2-6}), indicating that cross-correlations between the stocks build up progressively from the most general market-wide ones at short time scales to the more specific intra-industry ones at long time scales.

From the perspective of practical application of the $q$MST graphs, it is important to notice that, since their topology depends on $q$, the correlation structure of the market is different if one considers different fluctuation amplitudes. That is, with help of such graphs one can quickly realize how this structure looks like in the volatile periods and how it looks like in the more typical or quiet periods. It can allow one for a fine tuning of the investment portfolios, for instance, in order to avoid their composition of the stocks that tend to be more strongly correlated during volatile times, which increases the portfolio's risk, and which can remain unnoticed if one uses standard measures. For example, in the considered interval of time (1998-1999) for $q=2.0$ the shares of Corning were the most closely related to the General Electric ones, while for larger $q$s it occurs that they were even more strongly related to the AIG shares, so that a simultaneous investment in both these stocks was in fact more hazardous than it might be expected from an analysis of the MST for $q=2.0$ or an analysis using the standard MST approach.

\begin{figure}[t]
\begin{center}
\includegraphics[scale=0.1]{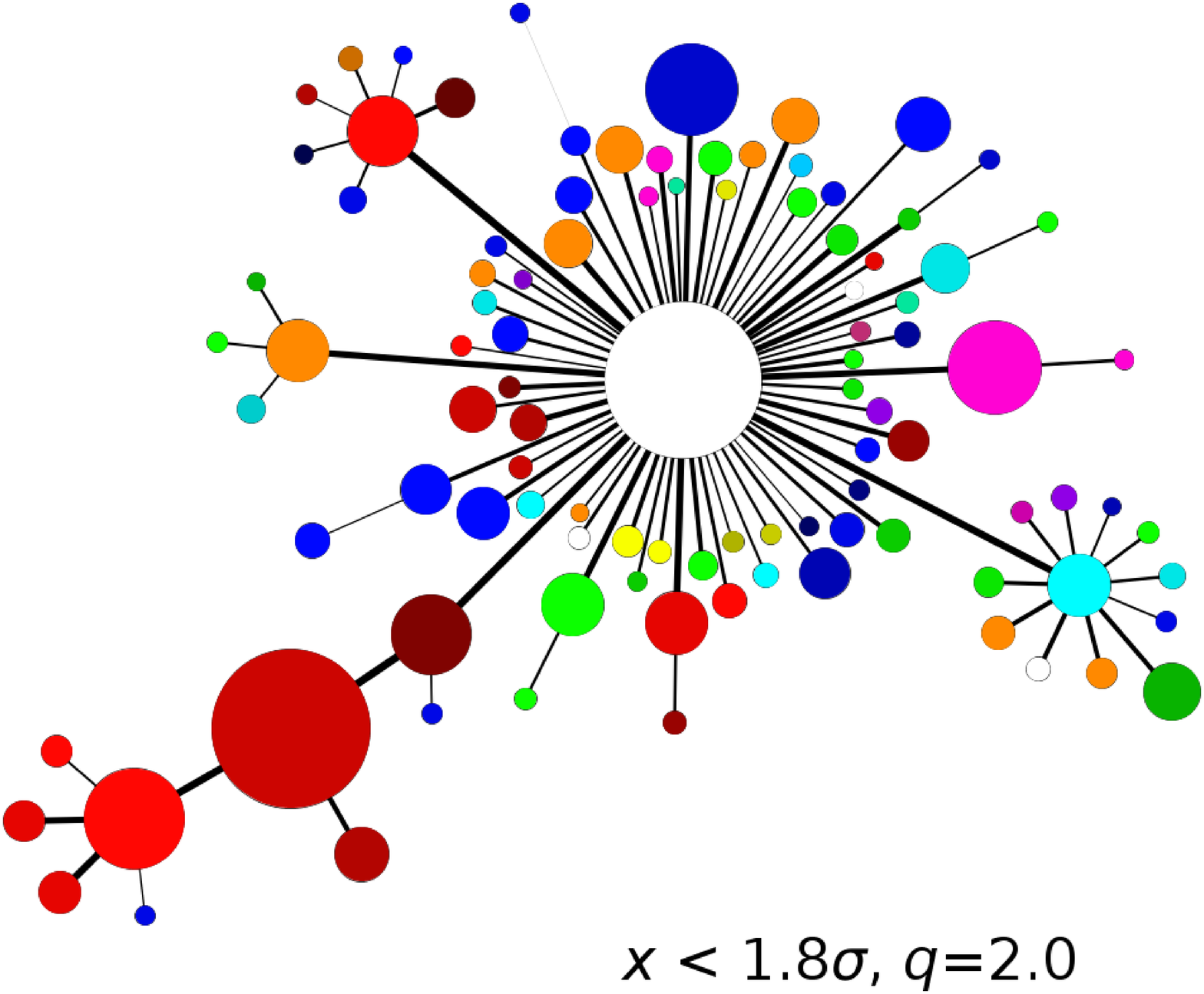}
\includegraphics[scale=0.1]{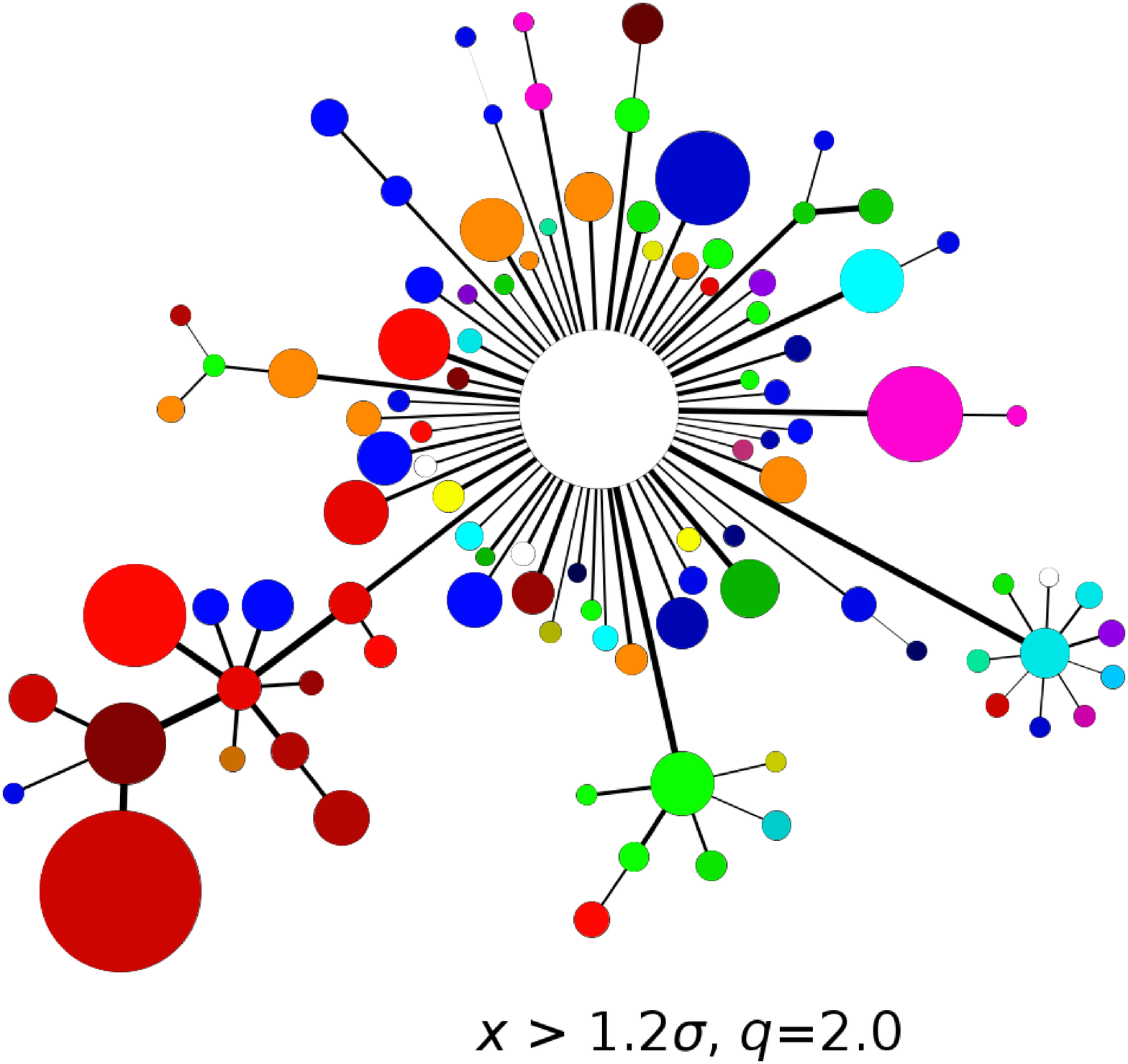}

\includegraphics[scale=0.1]{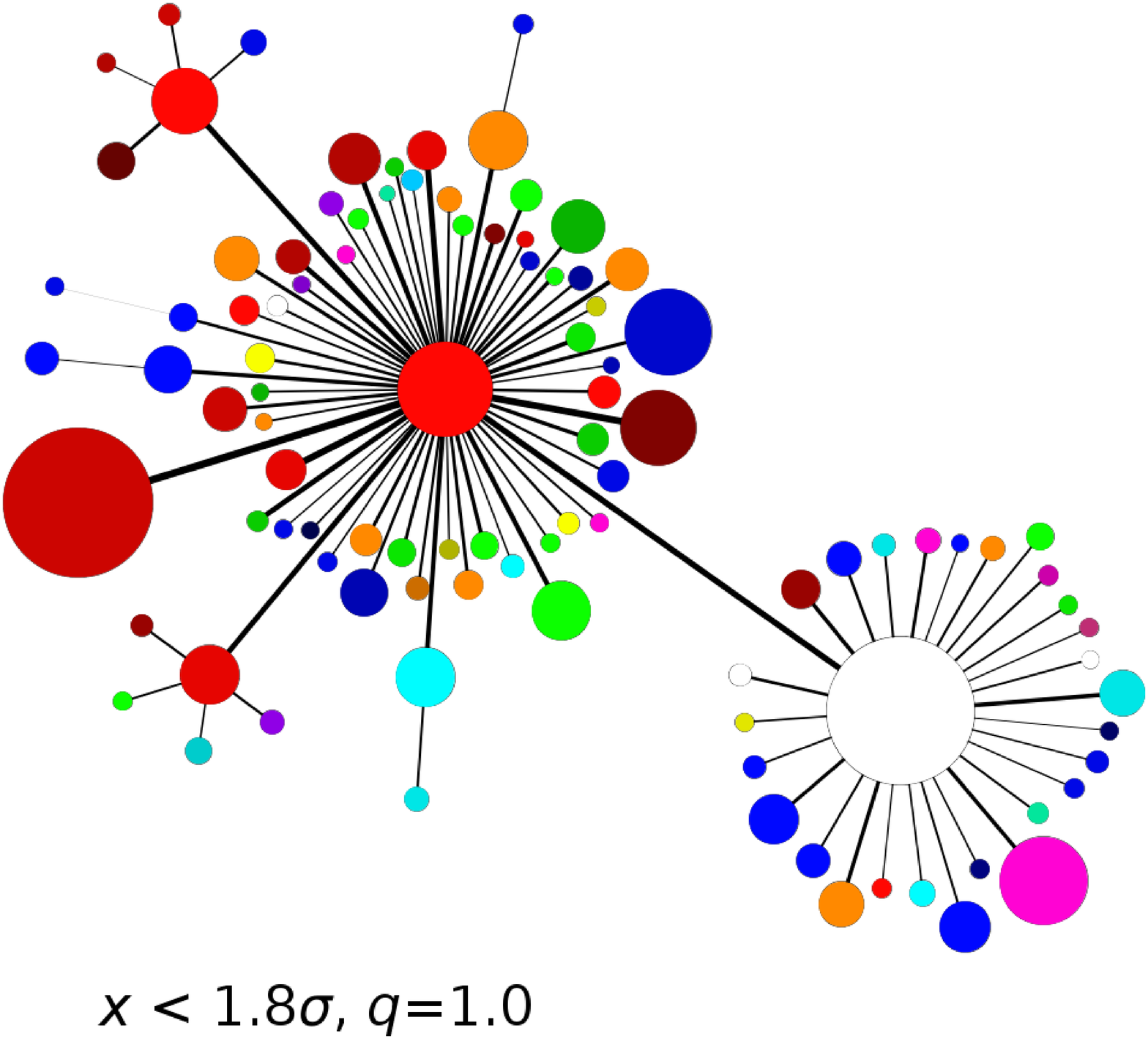}
\includegraphics[scale=0.1]{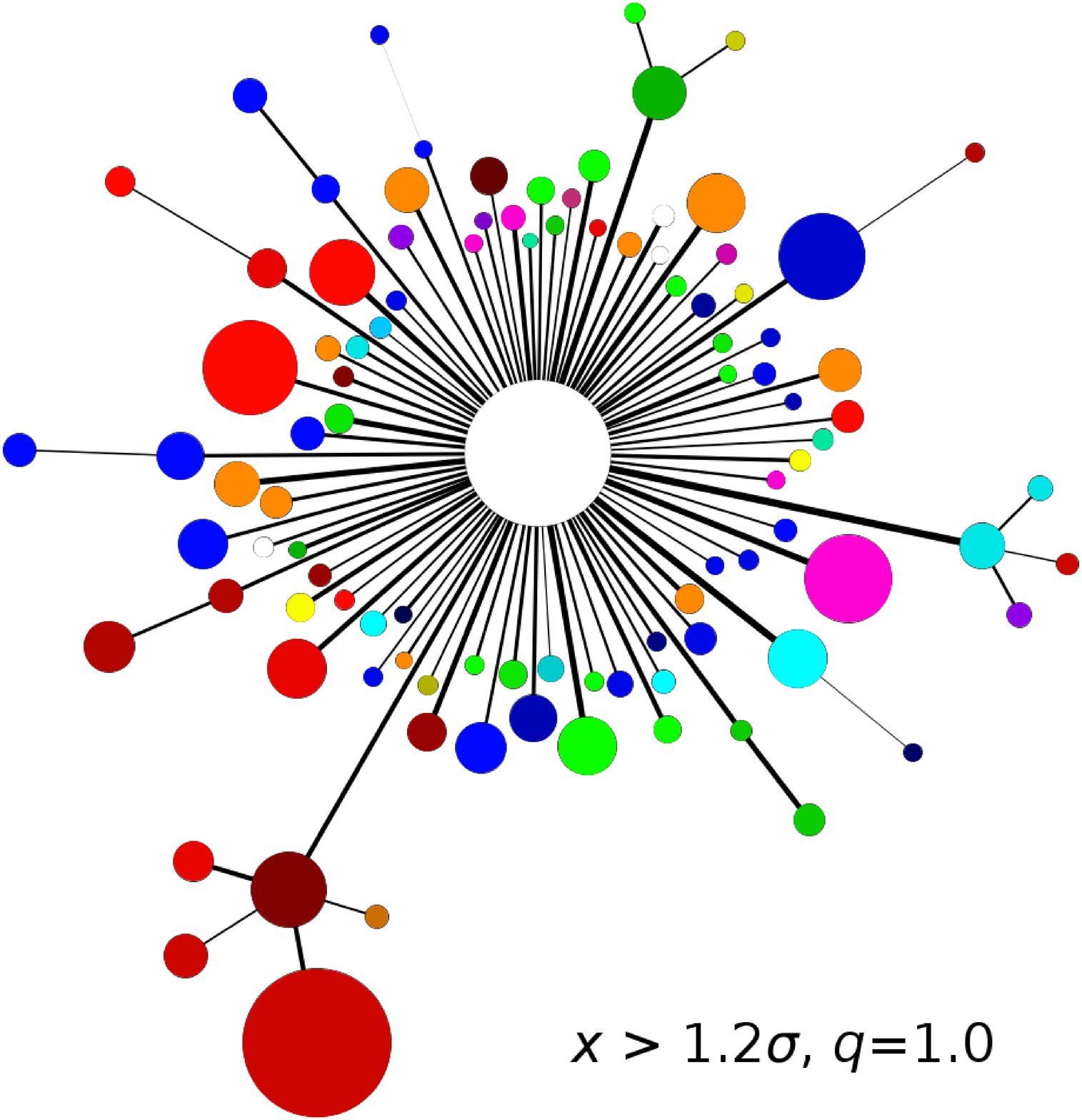}

\includegraphics[scale=0.08]{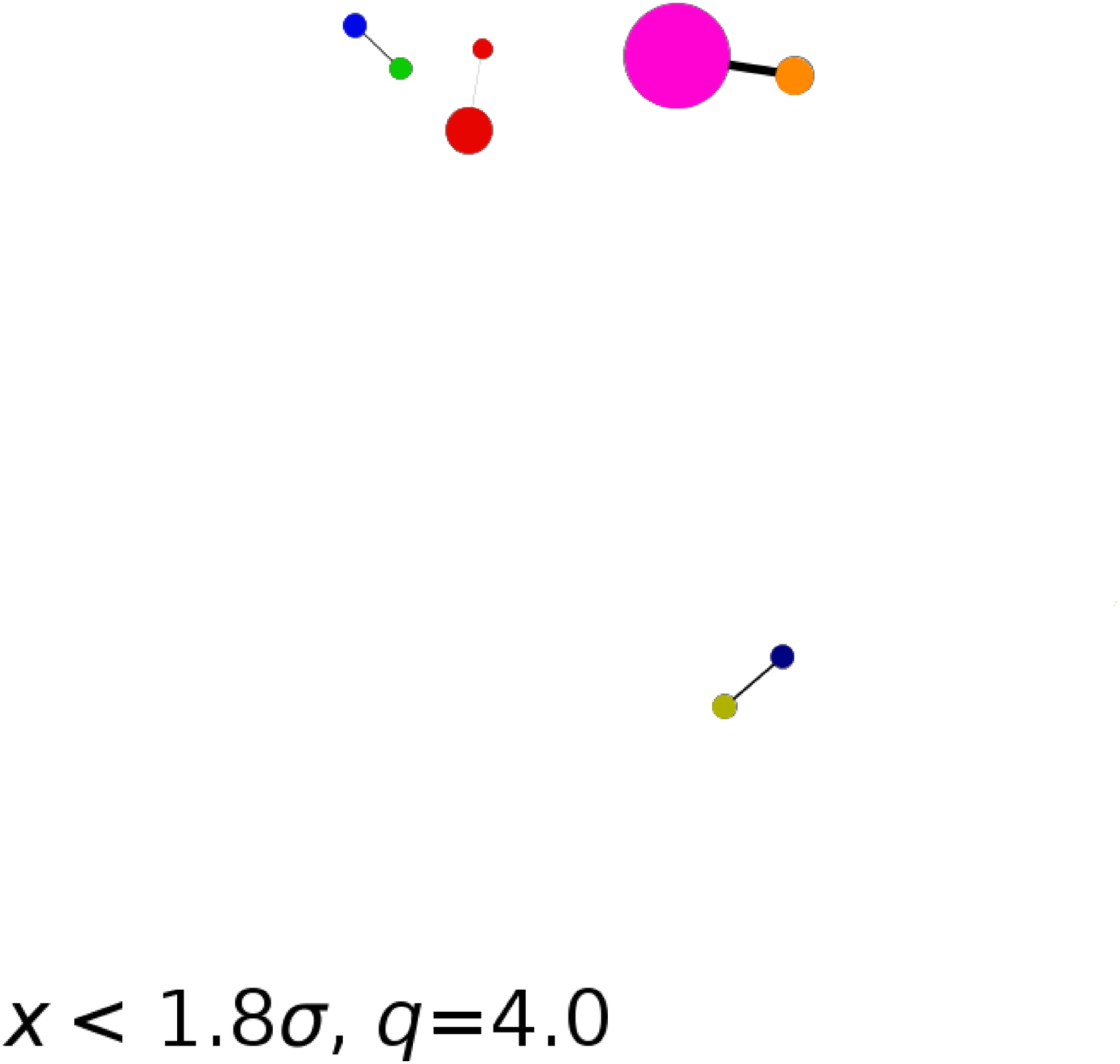}
\includegraphics[scale=0.08]{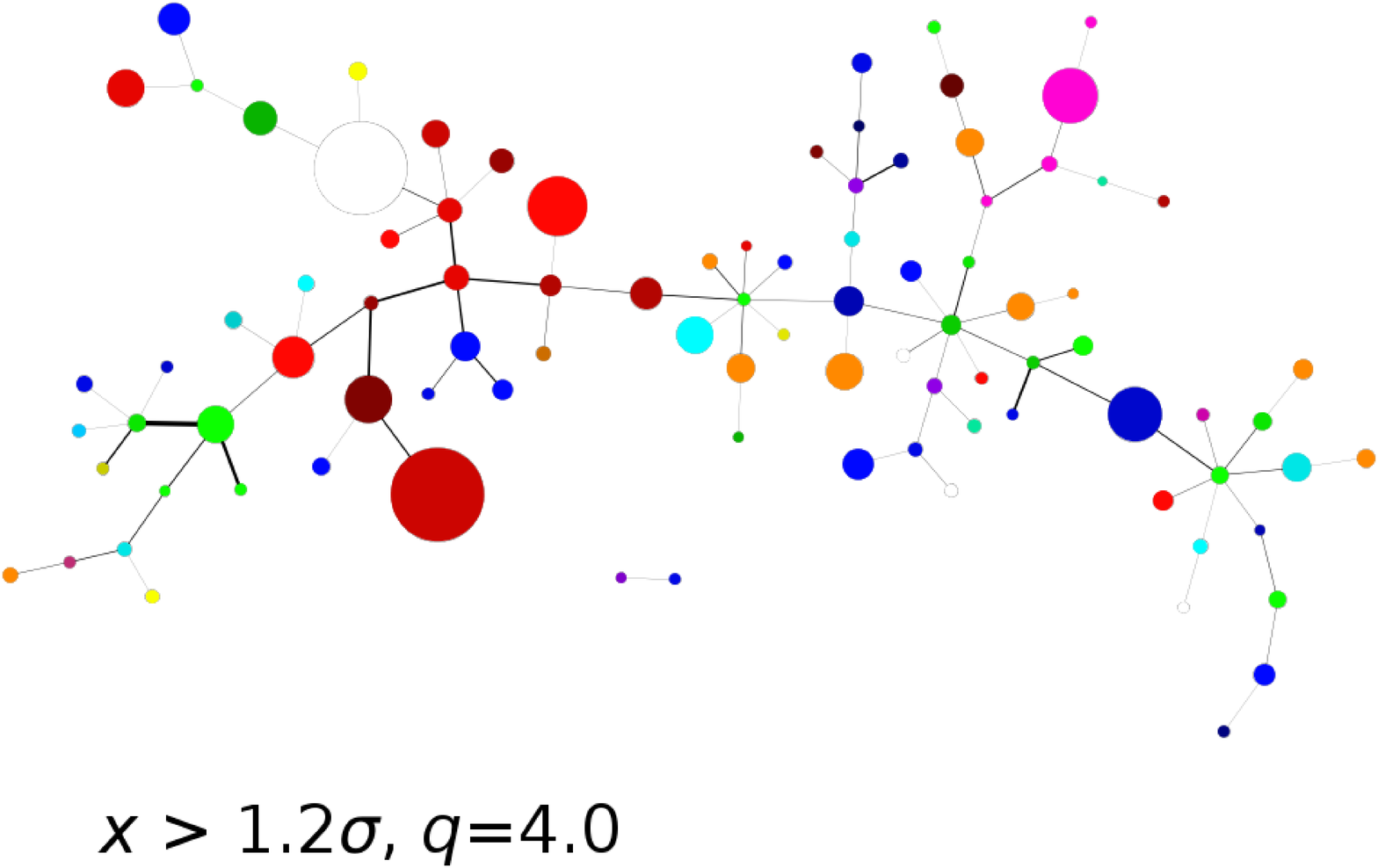}
\end{center}
\caption{(Color online) The $q$-dependent minimum spanning trees ($\Delta t=1$ min, $s=20$ min) created for two data sets with different statistical properties: Set 1 (large fluctuations randomized, small and medium ones preserved) on the left and Set 2 (small and medium fluctuations randomized, large fluctuations preserved) on the right. The respective $q$MSTs for different values of $q$ are compared: $q=2.0$ (top, $c_e = 51$ edges common to both sets), $q=1.0$ (middle, $c_e=21$), and $q=4.0$ (bottom, $c_e=0$). The trees were filtered to remove statistically insignificant edges ($\rho_q < \tau_{\rho}$) and nodes with degree $k=0$.}
\label{fig::discern}
\end{figure}

Another situation, in which the generalized $q$MST graphs are preferred over the $q=2$ MSTs, is illustrated in Fig.~\ref{fig::discern}. We consider two data sets that differ from each other in their statistical properties. Both data sets consist of the signals that were obtained from the original time series of returns ($\Delta t=1$ min, $s=20$ min) by partial randomization based on amplitude. Each signal was randomized independently. Set 1 comprises the signals in which large fluctuations ($r(t) > 1.8\sigma$) were shuffled among themselves, while the small and medium fluctuations ($r(t) \le 1.8\sigma$) were preserved as they occured in the signals ($\sigma$ stands for standard deviation). Set 2 was produced in similar manner, but here the small and medium fluctuations ($r(t) < 1.2\sigma$) were shuffled among themselves, while large ones ($r(t) \ge 1.2\sigma$) were preserved. Such randomization destroys all the correlations carried by the fluctuations of the respective amplitudes. Thus, only the correlations within the range ($1.2\sigma,1.8\sigma$) remained common to both sets, while the correlations outside that range were different in each set.

Despite that such differences exist, the $q$MSTs for $q=2$ reveal comparable topologies for both data sets as indicated in the top panels of Fig.~\ref{fig::discern}. To express this quantitatively, we calculated how many edges are common to both trees (by a common edge we mean an edge connecting the same pair of nodes in both cases): the respective number is $c_e=51$ for $q=2$. This indicates that the $q=2$ MSTs are not sufficiently sensitive to differentiate between the sets. In contrast, the differences appear more significant for both $q=1$ (middle panels, $c_e=21$ common edges) and $q=4$ (bottom panels, no common edges). In the latter case, the discrepancy is particularly striking as the tree for Set 1 is essentialy random and virtually disappears if tested against a null hypothesis of no genuine correlations, while almost the whole tree structure for Set 2 is preserved as being statistically significant. In fact, we expected this result, because for $q=4$ only those parts of the signals are taken into consideration that comprise many high-amplitude fluctuations in any pair of the signals and, after reshuffling, the large fluctuations become decorrelated in Set 1, while they are still correlated in Set 2. Therefore, in this situation, applying the generalized $q$MSTs proved advantageous since it allowed us to differentiate between the analyzed data sets.

\begin{figure}[t]
\begin{center}
\includegraphics[scale=0.09]{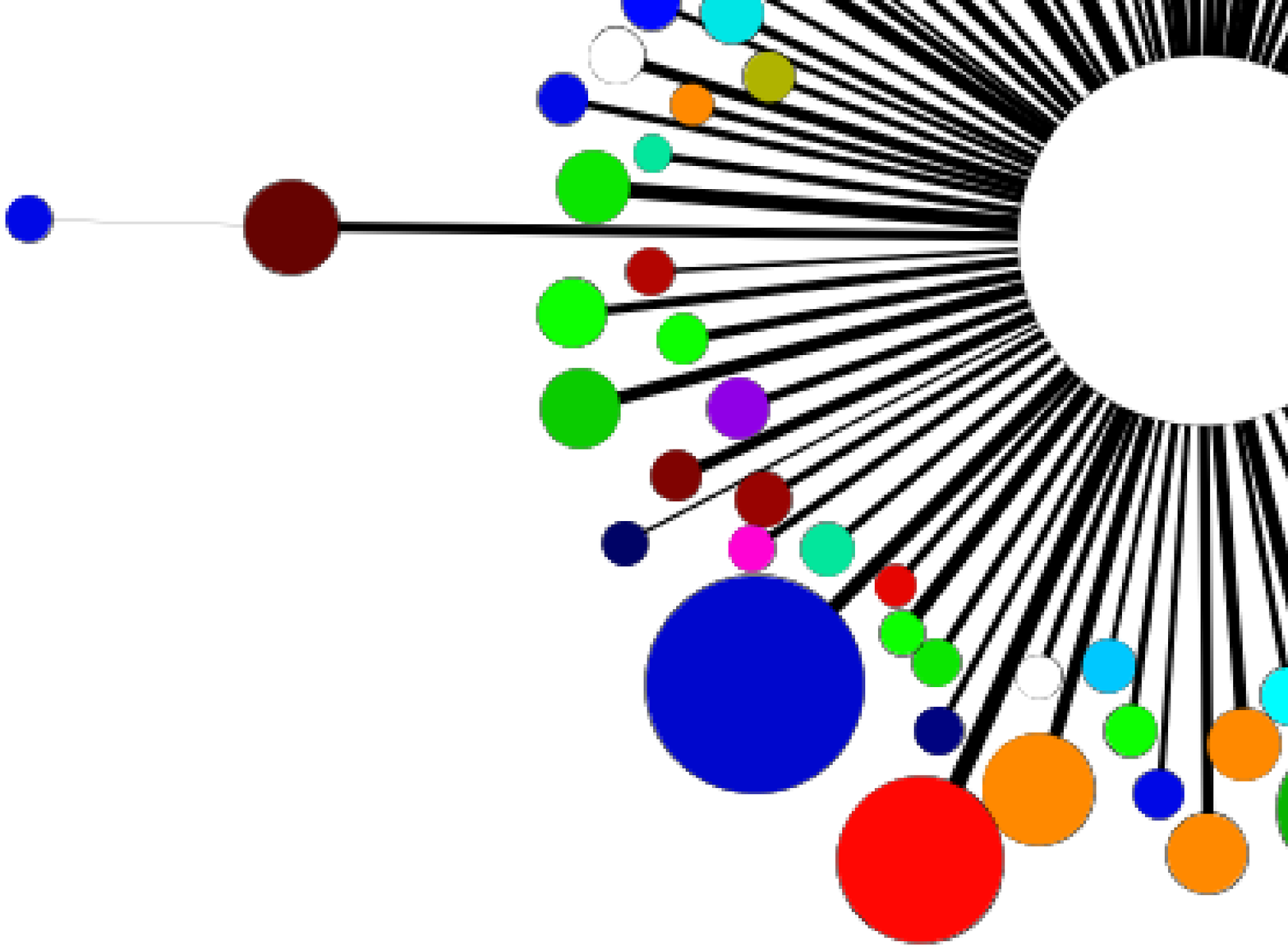}
\includegraphics[scale=0.1]{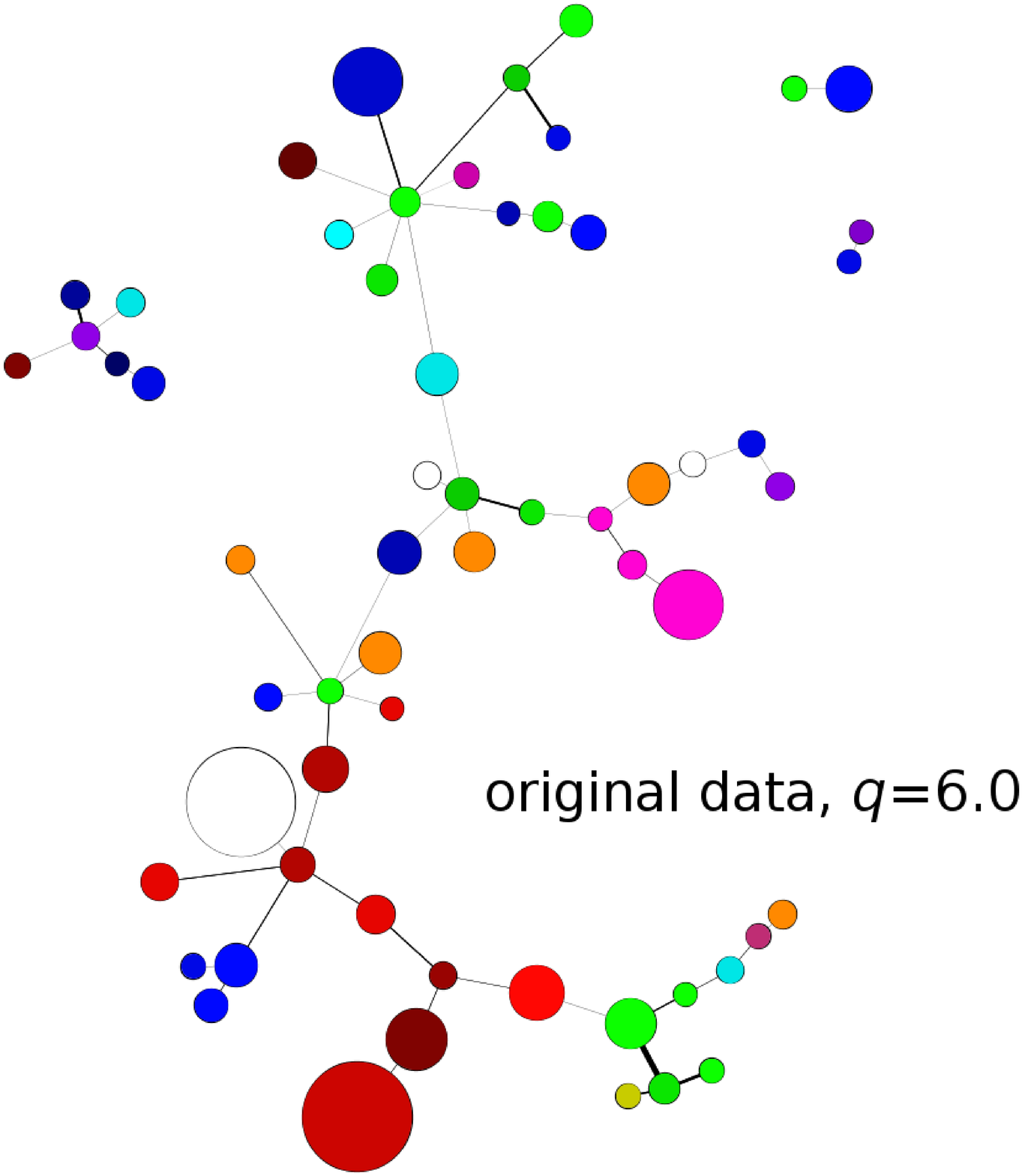}

\includegraphics[scale=0.09]{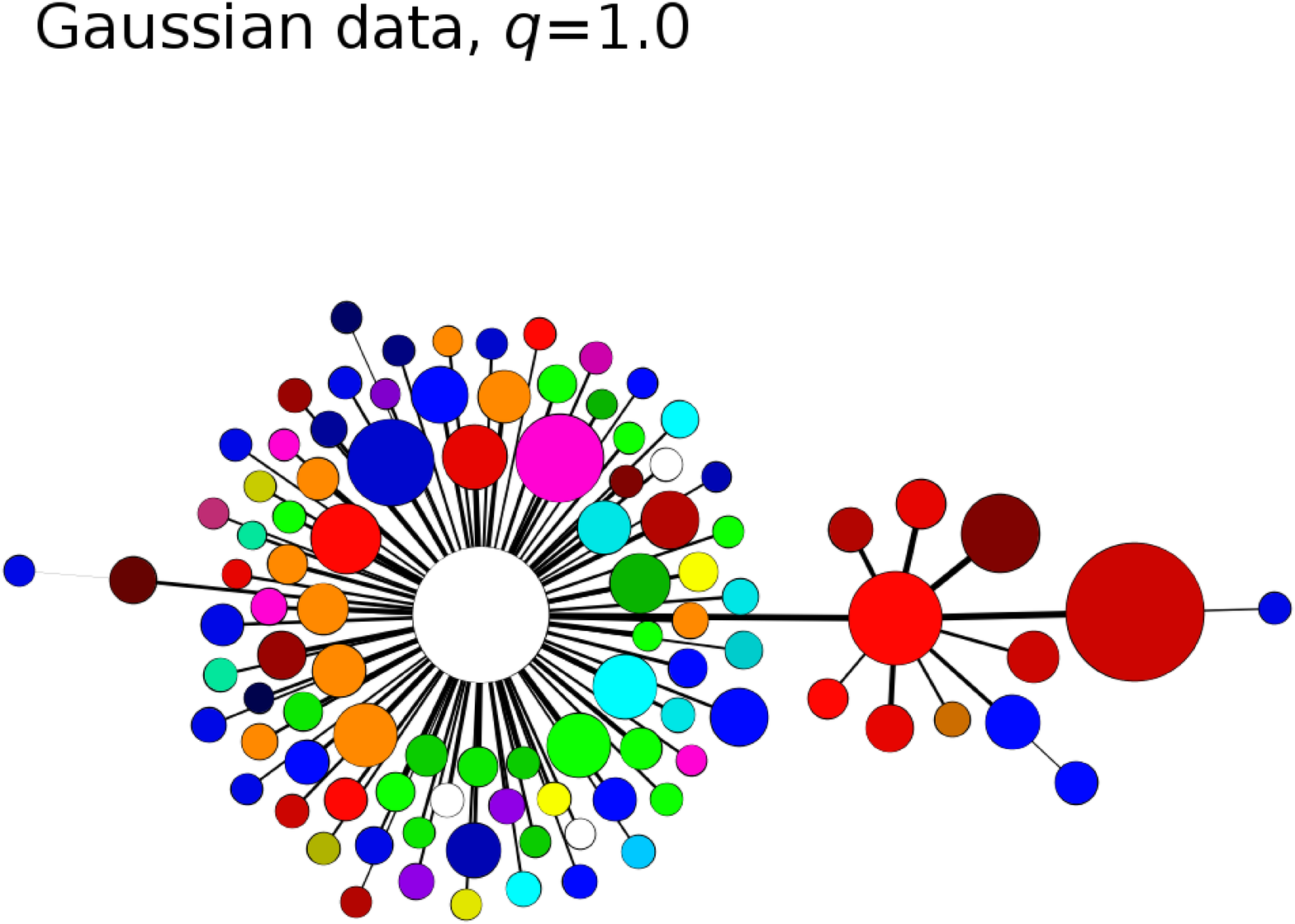}
\includegraphics[scale=0.08]{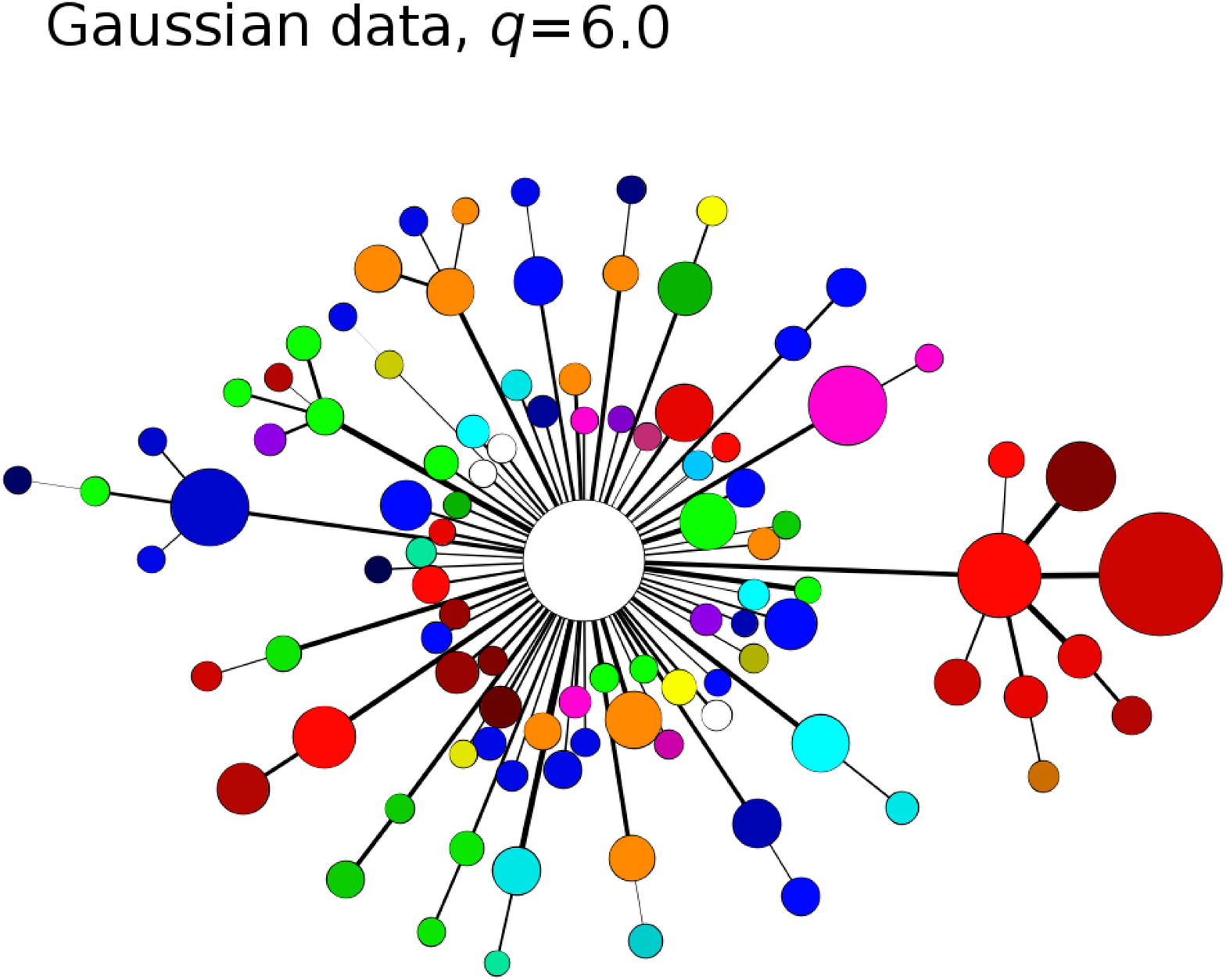}

\includegraphics[scale=0.08]{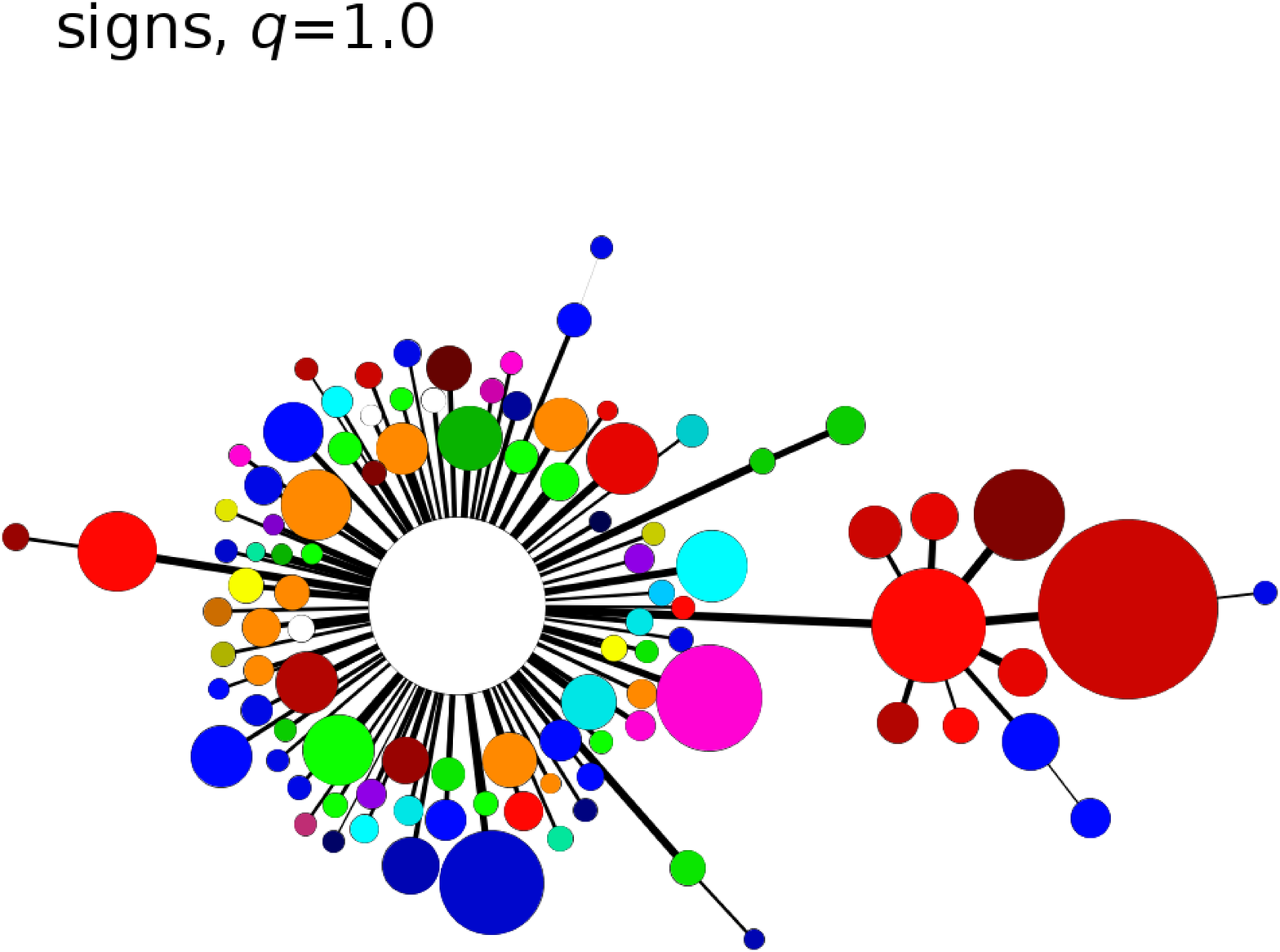}
\includegraphics[scale=0.1]{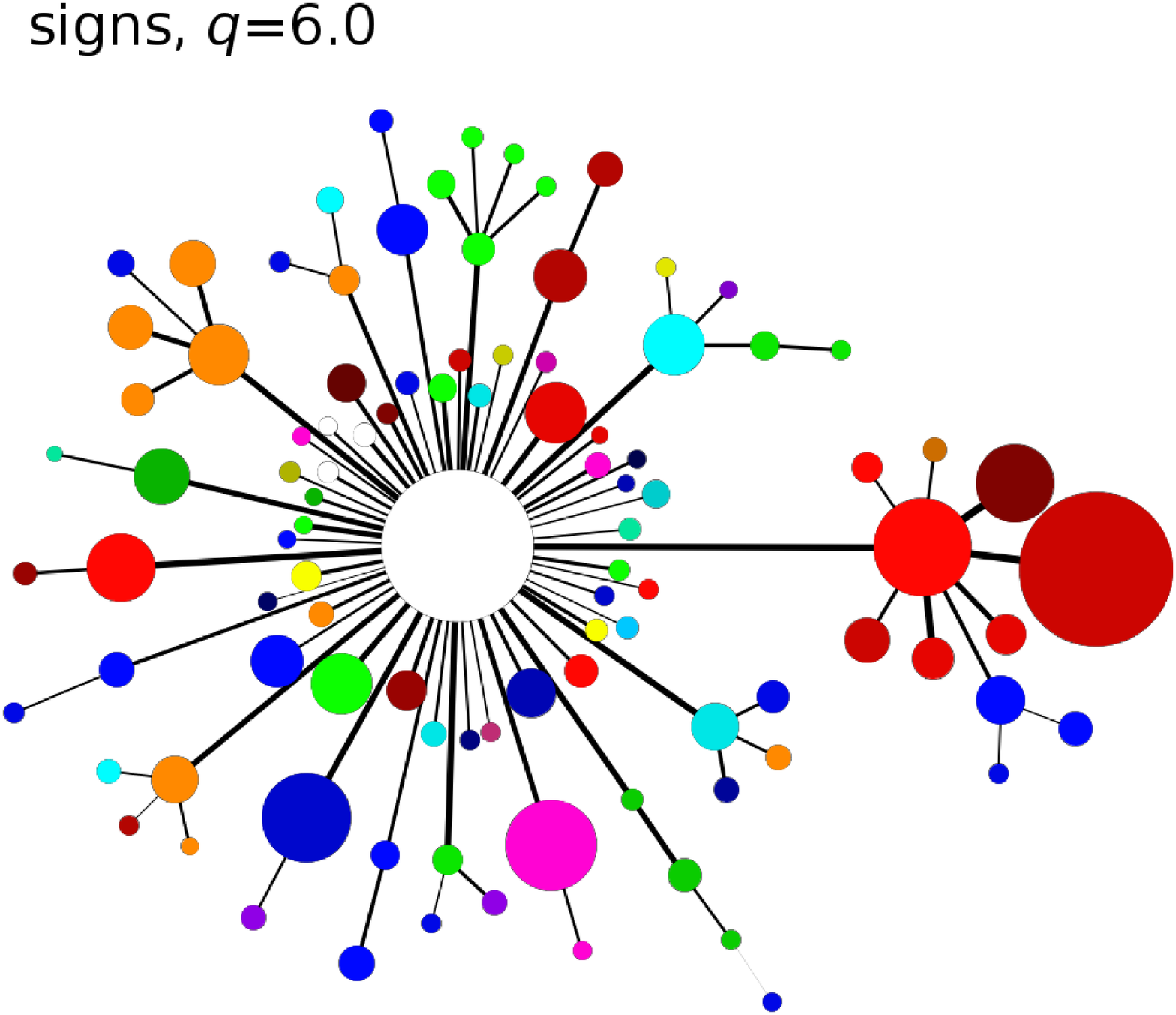}
\end{center}
\caption{(Color online) The $q$-dependent minimum spanning trees created for three data sets ($N=100$): the original time series of stock returns ($\Delta t=1$ min, $s=20$ min, top), the time series of Gaussianized returns (middle), and the time series of return signs (bottom). Two different values of $q$ are used: $q=1.0$ (left) and $q=6.0$ (right) and the number of edges common to the respective $q$MSTs is computed: $c_e=2$ (original returns), $c_e=71$ (Gaussian returns), and $c_e=65$ (return signs). The trees were filtered to remove statistically insignificant edges ($\rho_q < \tau_{\rho}$) and nodes with degree $k=0$.}
\label{fig::stability}
\end{figure}

Topological variability of the $q$MST graphs for a range of $q$ may also be considered a property of data, expressing stability of the correlation structure for different fluctuation amplitudes. The more static the topology of $q$MSTs for different $q$s is, the broader range of fluctuations is correlated in the same manner. Fig.~\ref{fig::stability} presents such an application of the $q$MST graphs. The trees calculated for the original set of time series of returns (top panels) are compared with the trees calculated for time series of the Gaussian returns (middle panels) and time series of the return signs (bottom panels). While the original returns have p.d.f. with the power-law tails with a scaling index $\alpha \approx -3$, the Gaussian returns were obtained from the original ones by suppressing their amplitudes in such a way that their p.d.f. became Gaussian, while the amplitude ranks of the fluctuations remained preserved~\cite{theiler1992}. The sign series were obtained by replacing the returns by their signs, i.e., $r'(t)=1$ if $r(t) > 0$, $r'(t)=-1$ if $r(t) < 0$, and $r'(t)=0)$ otherwise. For a larger effect, in Fig.~\ref{fig::stability} we compare $q$MSTs for $q=1$ (left) and $q=6$ (right). As the most variable (i.e., different for different stock pairs) correlations are observed in the original data for the largest fluctuations (only $c_e=2$ edges are common to $q=1$ and $q=6$), suppressing these fluctuations by either Gaussianizing their p.d.f. or by considering only their signs leads to more uniform strength of the correlations ($c_e=71$ and $c_e=65$ common edges, respectively). Thus, $q$MSTs presented in the top panels of Fig.~\ref{fig::stability} have much more heterogeneous edge weights than their counterparts shown in the middle and bottom panels. 

In this context, one can view the parameter $q$ in $q$MST in the same way as it is commonly viewed in the studies using the R\'enyi entropy or in the multifractal analysis, where one computes a spectrum of the generalized fractal dimensions $D_q$, the generalized Hurst exponents $H_q$, or the multifractal spectrum $\tau(q)$~\cite{schuster2006}. By applying different values of $q$, one selects certain parts of analyzed data and acquires some insight into their properties. Here, by varying $q$, we obtain a family of trees describing correlations between different parts of signals based on standard deviations of their fluctuations. This makes the $q$MST graphs be a possible method of quantifying heterogeneity of correlations in a way resembling the way the singularity spectra $f(\alpha)$ quantify the richness of multifractal structures.

\begin{figure}[t]
\begin{center}
\includegraphics[scale=0.08]{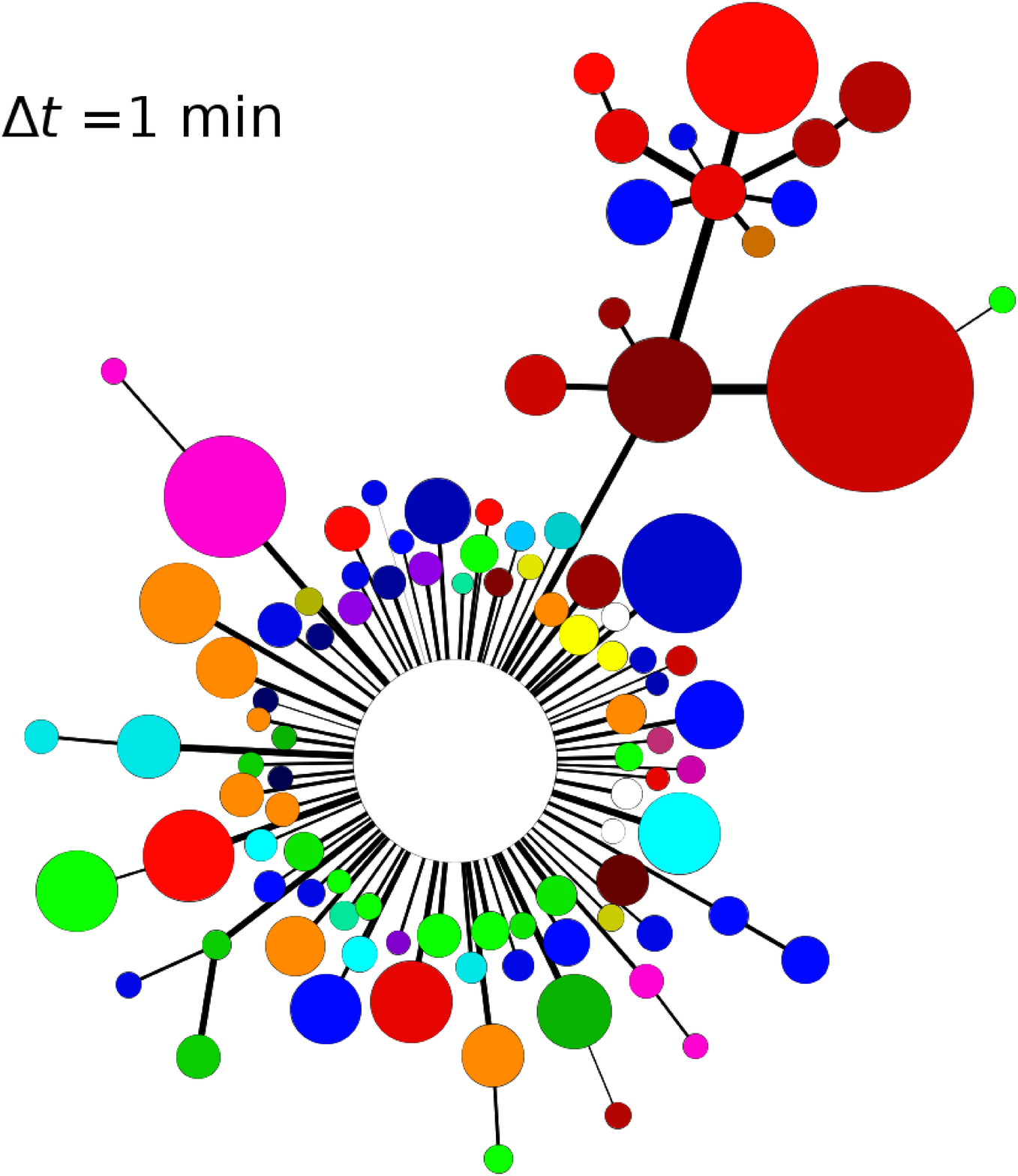}
\includegraphics[scale=0.08]{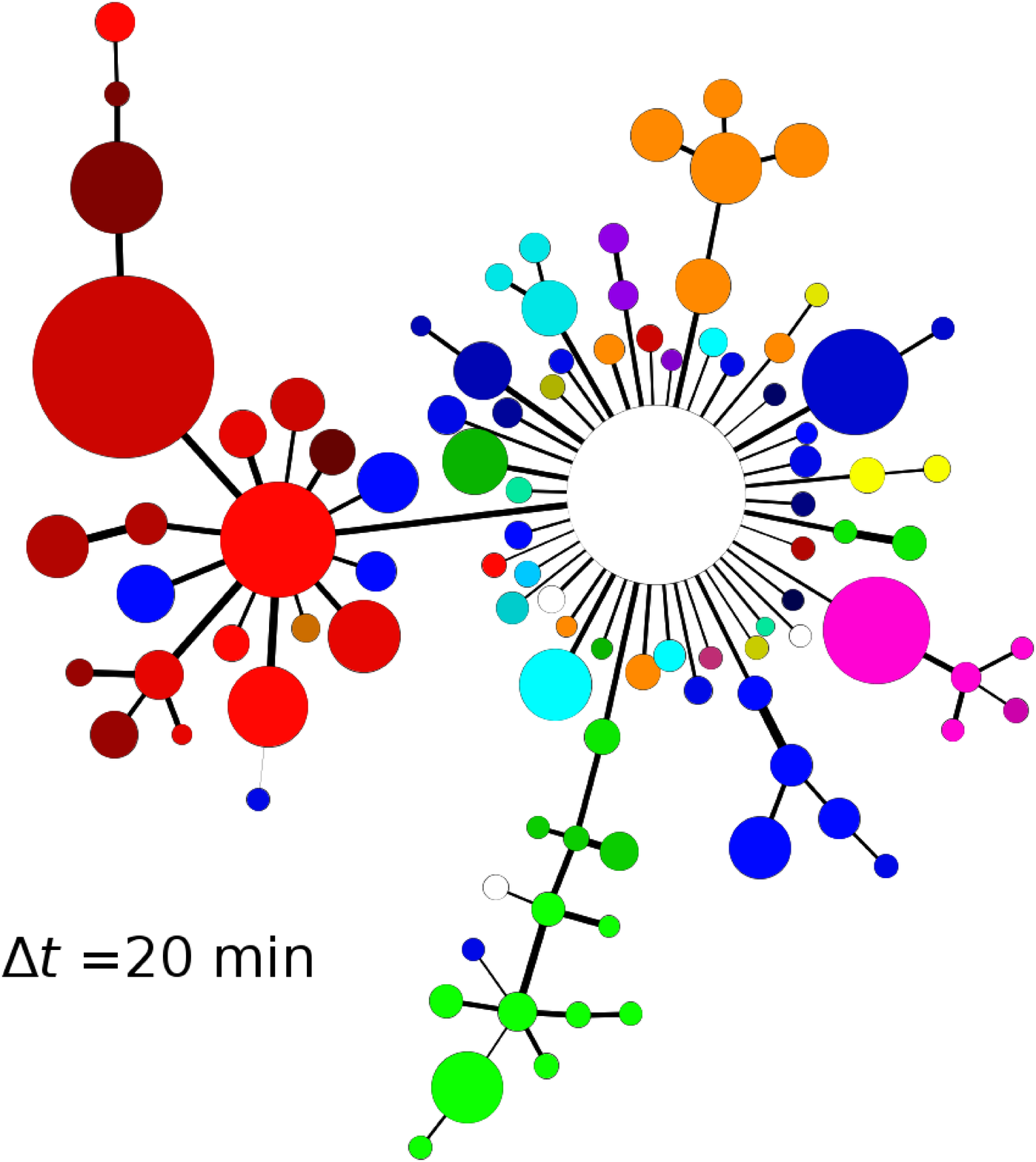}

\includegraphics[scale=0.08]{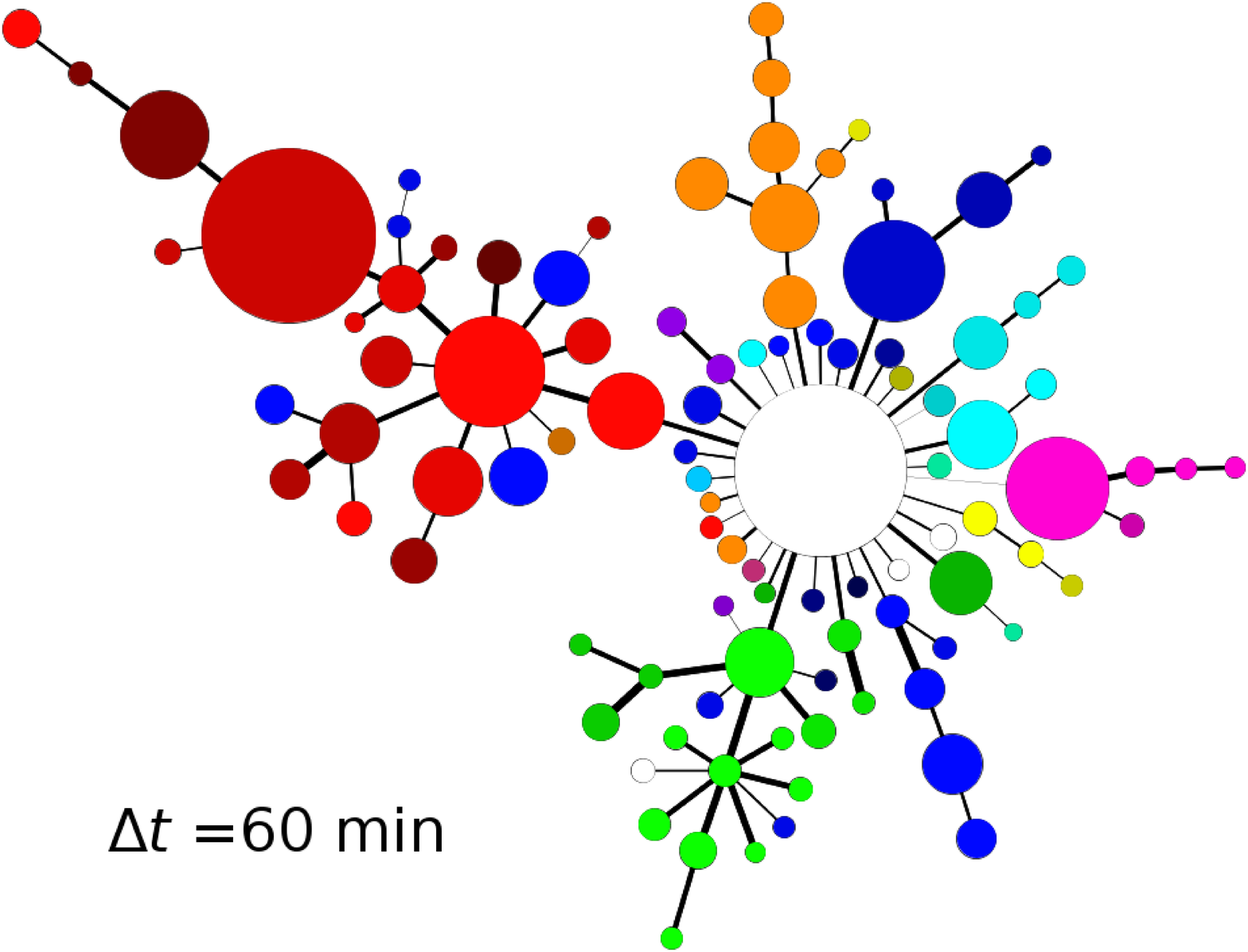}
\includegraphics[scale=0.08]{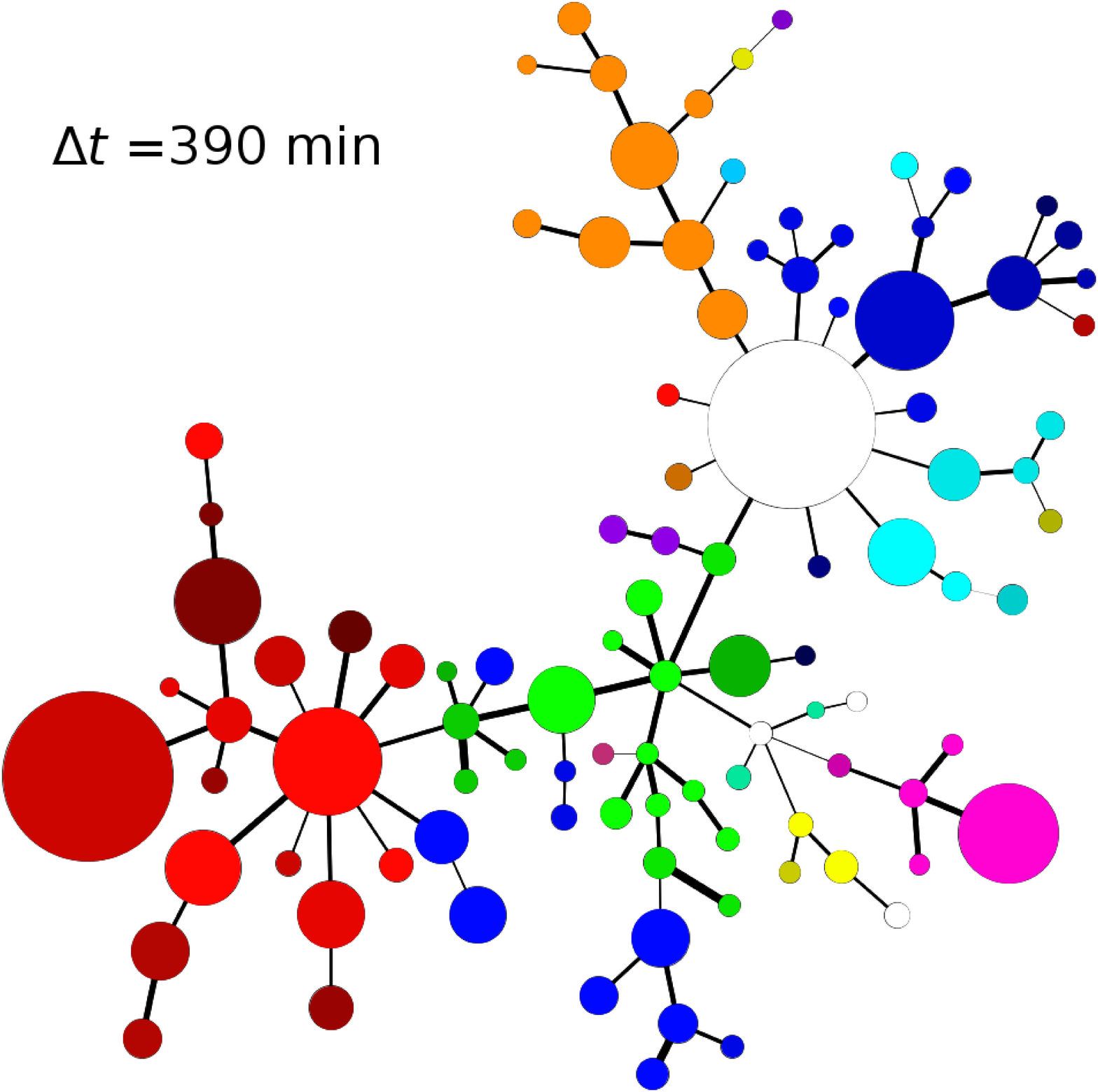}
\end{center}
\caption{(Color online) The minimum spanning trees constructed in the standard way by using the Pearson correlation coefficient for time series of returns with $\Delta t=1$ min. (top left), 20 min. (top right), 60 min. (bottom left), and 390 min. (bottom right).}
\label{fig::standard.msts}
\end{figure}

For the sake of comparison, we also present the minimal spanning trees calculated in accordance with their standard definition, i.e., by using the Pearson correlation coefficients~\cite{mantegna1999} for the non-detrended time series of returns. A straightforward choice of the data sampling interval could be the same $\Delta t=1$ min. as before, but one has to realize that there is not any direct counterpart of the temporal scale $s$ in this case. The difference is that in the DFA-related approaches, the temporal scale is determined by a window size, in which detrending is performed and in which variance of the residual signal is calculated, while in the standard approach the time scale is associated directly with the sampling interval $\Delta t$. Therefore, here we have to check different values of $\Delta t$: 1, 20, 60, and 390 min. and to construct one MST for each of them. Fig.~\ref{fig::standard.msts} shows the resulting trees. As usual in such analyses, by increasing the time scale, we obtain more and more visible clusters that may be identified with the market sectors. By comparing the present trees with the ones in Fig.~\ref{fig::s20.q1-6}, Fig.~\ref{fig::s60.s390.q2-6}, and Fig.~\ref{fig::s1950.s7800.q2-6}, it is evident that they resemble more the centralized trees for $1.0 \le q \le 3.0$ than the trees for $q \ge 4.0$ that were significantly less centralized.

In order to compare the results of both methods in a quantitative manner, we prefer to consider the matrices consisting of all the respective coefficients for all $N(N-1)/2=4950$ pairs of stocks instead of considering only the minimal spanning trees. This is because by taking all possible coefficients into consideration, we will increase a statistical significance of our comparison. Therefore, here we take the values of $\rho_q$ and the Pearson correlation coefficients $c$ for all pairs of signals and investigate how similar their sets are for different choices of the parameters: $\Delta t$, $s$, and $q$. For each choice, we order both coefficients into vectors in such a way that at first we take $N-1$ coefficients (of the same type) involving a stock $X_1$, then the $N-2$ coefficients involving a stock $X_2$, then $X-3$ coefficients involving a stock $X_3$, an so on, and finally the coefficient for stocks $X_{N-1}$ and $X_N$.

By proceeding along this way, we eventually obtain two vectors: ${\bf c}(\Delta t)$ and $\bm{\rho}_q(s)$ for each $\Delta t$, $q$, and $s$. (Actually, the coefficient $\rho_q(s)$ also depends on some sampling time $\Delta t'$, but as we do not change the latter and consequently use the data sampled with $\Delta t'=1$ min., we neglect $\rho_q(s)$ dependence on $\Delta t'$). Then we calculate a normalized scalar product of these vectors:
\begin{eqnarray}
P(\Delta t,s,q)= {\sum_{m=1}^{4950} c_m(\Delta t) \rho_q^{(m)}(s) \over |{\bf c}(\Delta t)| |\bm{\rho}_q(s)|}, \\ -1 \le P(\Delta t,s,q) \le 1,
\nonumber
\label{eq::scalar}
\end{eqnarray}
where the vector components labeled by $m$ correspond to unique stock pairs ($m=1,...,4950$). We prefer the normalized scalar product over the Pearson  correlation coefficient here, because we do not want the matrix elements $c_{X_i X_j}$ and $\rho_q^{(X_i X_j)}$ to be altered by the normalization.

\begin{table}
\begin{tabular}{@{}|c||c||c|c|c|c|c|c|@{}}
\hline
$\downarrow$ $\Delta t$ & {\backslashbox{$s$}{$q$}} & 1.0 & 2.0 & 3.0 & 4.0 & 5.0 & 6.0 \\ \hline\hline
1 & 20 & {\bf 0.97} & {\bf 0.98} & 0.92 & 0.61 & 0.33 & 0.20 \\ \cline{2-8}
 & 60 & {\bf 0.96} & {\bf 0.96} & 0.95 & 0.86 & 0.70 & 0.53 \\ \cline{2-8}
 & 390 & 0.95 & 0.95 & 0.94 & 0.90 & 0.84 & 0.77 \\ \cline{2-8}
 & 7800 & 0.89 & 0.87 & 0.83 & 0.77 & 0.71 & 0.66 \\ \hline\hline

20 & 20 & {\bf 0.96} & {\bf 0.95} & 0.86 & 0.53 & 0.28 & 0.17 \\ \cline{2-8}
 & 60 & {\bf 0.97} & {\bf 0.97} & 0.95 & 0.84 & 0.68 & 0.52 \\ \cline{2-8}
 & 390 & {\bf 0.98} & {\bf 0.98} & {\bf 0.96} & 0.92 & 0.86 & 0.79 \\ \cline{2-8}
 & 7800 & 0.94 & 0.92 & 0.88 & 0.82 & 0.77 & 0.71 \\ \hline\hline

60 & 20 & {\bf 0.95} & 0.95 & 0.86 & 0.55 & 0.29 & 0.19 \\ \cline{2-8}
 & 60 & {\bf 0.97} & {\bf 0.97} & 0.95 & 0.85 & 0.69 & 0.53 \\ \cline{2-8}
 & 390 & {\bf 0.98} & {\bf 0.99} & {\bf 0.98} & 0.94 & 0.88 & 0.81 \\ \cline{2-8}
 & 7800 & {\bf 0.97} & {\bf 0.96} & 0.93 & 0.89 & 0.83 & 0.78 \\ \hline\hline

390 & 20 & 0.91 & 0.90 & 0.83 & 0.53 & 0.28 & 0.19 \\ \cline{2-8}
 & 60 & 0.93 & 0.93 & 0.91 & 0.83 & 0.70 & 0.53 \\ \cline{2-8}
 & 390 & {\bf 0.95} & {\bf 0.95} & 0.95 & 0.92 & 0.86 & 0.79 \\ \cline{2-8}
 & 7800 & {\bf 0.98} & {\bf 0.98} & {\bf 0.96} & 0.91 & 0.86 & 0.81 \\ \hline
\end{tabular}
\caption{Values of the normalized scalar product for 4 Pearson-coefficient-based ($\Delta t=1,20,60,390$ min.) and 24 $\rho_q$-based ($s=20,60,390,7800$ min., $q=1.0,2.0,3.0,4.0,5.0,6.0$) weighted networks with nodes representing the 100 largest companies. The values greater than 0.950 are distinguished in bold.}
\label{tab::scalar}
\end{table}

The results are collected in Tab.~\ref{tab::scalar}. They show that both methods coincide most for $1 \le q \le 2$ (or sometimes up to $q=3$) and for $s \ge \Delta t$ (but not for $s \gg \Delta t$), exactly as one would expect on formal ground. In this case $q$MSTs offer similar information as the standard MST. In contrast, if $s$ is of order of an hour or less, for $q \ge 4$ the presently introduced method provide us with information that cannot be found with the standard approach. This means that in the latter case, by using the detrended minimal spanning trees of order $q$, we obtain significantly extended information about the correlation structure of the data under study. This constitutes a strong quantitative argument in favor of the new method.

\section{Conclusions}
\label{sect.4}

In this work we introduced a family of $q$-dependent minimum spanning trees that are able to visualize the cross-correlations that are restricted to a specific range of the fluctuation amplitudes. These trees are a generalization of the detrended MSTs proposed in~\cite{wang2013a}. They are defined on the basis of the $q$-dependent detrended cross-correlation coefficient $\rho_q$ by transforming it to a metric distance. We applied the technique of the $q$MST graphs to visualize the correlation structure of the American stock market and found that by changing the fluctuation amplitudes and time scales we are focused on, one observes substantial changes in topology of the graphs. We identified a few previously known features of the financial data like the Epps effect, the division of the market into industrial sectors, presence of the so-called market factor, and the priviledged role of some specific stocks (like General Electric and Cisco) in the market structure. We showed that the trees for small values of $q$ (e.g., $q \le 3$) exhibit clusters that are strongly related to the industrial sectors of the market, especially for longer temporal scales $s$, while for larger values of this parameter ($q \ge 4$) such industrial structure becomes less and less evident. This indicates thus that what binds these industrial clusters are the typical fluctuations of moderate amplitude, while the largest fluctuations are more likely to synchronize among the stocks that do not neccessarily belong to the same industries. 

The detrended $q$MSTs can complement the coefficient $\rho_q$ in disentangling ``hidden'' cross-correlations that can be observed neither in analyses based on the measures defined for $q=2.0$, like $\rho_{\rm DCCA}$ and the related detrended MSTs, nor in analyses using the standard minimal spanning trees. In principle, the role of MST is to decrease significantly the number of edges in a network by focusing on the most important ones. By considering a family of such trees with varying $q$, the number of edges remains the same in each tree but the routes of connections among nodes may change with different choices of $q$. Changes indicate heterogeneity of cross-correlations, like, for instance, their varying strength for different fluctuation amplitude in the time series, while independence of $q$ provides evidence for homogeneity of cross-correlations. Our generalisation of the MST concept to $q$MST indicates a direction of studying and quantifying such effects.

In order to illustrate how advantageous is the $q$MST approach, we applied it to two data sets with different, known statistical properties and documented that the trees representing $q=2$ performed poor in distinguishing between these sets, unlike the generalized trees with $q \ne 2$, whose performance was satisfactory. We thus showed that $q$MSTs, if constructed for different choices of $q$, are capable, as expected, to detect the diversity of correlations. Topological variability of the trees for different $q$s is related to how diverse the correlations are. Therefore, we envisage that both $\rho_q$ and $q$MSTs will prove helpful in portfolio analysis and, of course, also in many other areas where cross-correlations are to be extracted from nonstationary time series.


\begin{thebibliography}{99}

\bibitem{barrow1985} J.D.~Barrow, S.P.~Bhavsar, D.H.~Sonoda, Minimal spanning trees, filaments and galaxy clustering, Month. Not. R.~Astron. Soc. {\bf 216}, 17-35 (1985).
\bibitem{adami1999} C.~Adami, A.~Mazure, The use of minimal spanning tree to characterize the 2d cluster galaxy distribution, Astron. Astrophys. Suppl. Ser. {\bf 134}, 393–400 (1999).
\bibitem{mantegna1999} R.N.~Mantegna, Hierarchical structure in financial markets, Eur. Phys.~J.~B {\bf 11}, 193-196 (1999).
\bibitem{zivkovic2006} J.~\u Zivković, B.~Tadić, N.~Wick, S.~Thurner, Statistical indicators of collective behavior and functional clusters in gene networks of yeast, Eur. Phys.~J.~B {\bf 50}, 255-258 (2006).
\bibitem{hernandez2007} E.~Hernandez-Garc\'{i}a, E.A.~Herrada, A.F.~Rozenfeld, C.J.~Tessone, V.M.~Egu\'{i}luz, C.M.~Duarte, S.~Arnaud-Haond, E.~Serr\~ao, Evolutionary and ecological trees and networks, Nonequilibrium Statistical Mechanics and Nonlinear Physics, Ed. by O. Descalzi, O.A. Rosso, H.A. Larrondo, AIP Conf. Proc. {\bf 913}, 78-83 (2007).
\bibitem{vandewalle2001} N.~Vandewalle, F.~Brisbois, X.~Tordoir, Non-random topology of stock markets, Quantit.~Fin. {\bf 1}, 373-374 (2001).
\bibitem{onnela2003} J.-P.~Onnela, A.~Chakraborti, K.~Kaski, J.~Kertesz, Dynamic asset trees and Black Monday, Physica A {\bf 324}, 247-252 (2003).
\bibitem{micciche2003} S.~Micciche, G.~Bonanno, F.~Lillo, R.N.~Mantegna, Degree stability of a minimum spanning tree of price return and volatility, Physica A {\bf 324}, 66-73 (2003).
\bibitem{bonanno2004} G.~Bonanno, G.~Caldarelli, F.~Lillo, C.~Miccich\`e, N.~Vandewalle, R.N.~Mantegna, Networks of equities in financial markets, Eur. Phys.~J.~E {\bf 38}, 363-371 (2004).
\bibitem{mcdonald2005} M.~McDonald, O.~Suleman, S.~Williams, S.~Howison, N.F.~Johnson, Detecting a currency's dominance or dependence using foreign exchange network trees, Phys. Rev.~E {\bf 72}, 046106 (2005).
\bibitem{mizuno2006} T.~Mizuno, H.~Takayasu, M.~Takayasu, Correlation networks among currencies, Physica A {\bf 364}, 336-342 (2006).
\bibitem{eom2007} C.Eom, G.~Oh, S.~Kim, Topological properties of a minimal spanning tree in the Korean and the American stock markets, J.~Kor. Phys. Soc. {\bf 51}, 1432-1436 (2007).
\bibitem{naylor2007} M.J.~Naylor, L.C.~Rose, B.J.~Moyle, Topology of foreign exchange markets using hierarchical structure methods, Physica A {\bf 389}, 199-208 (2007).
\bibitem{coelho2007} R.~Coelho, C.G.~Gilmore, B.~Lucey, P.~Richmond, S.~Hutzler, The evolution of interdependence in world equity markets - evidence from minimum spanning trees, Physica A {\bf 376}, 455-466 (2007).
\bibitem{gorski2008} A.Z.~Górski, S.~Drożdż, J.~Kwapień, Scale free effects in world currency exchange network, Eur. Phys.~J.~B {\bf 66}, 91-96, (2008).
\bibitem{garas2008} A.~Garas, P.~Argyrakis, S.~Havlin, The structural role of weak and strong links in a financial market network, Eur.~Phys.~J.~B {\bf 63}, 265-271 (2008).
\bibitem{sieczka2009} P.~Sieczka, J.A.~Hołyst, Correlations in commodity markets, Physica A {\bf 388}, 1621-1630 (2009).
\bibitem{kwapien2009} J.~Kwapień, S.~Gworek, S.~Drożdż, A.Z.~Górski, Analysis of a network structure of the foreign currency exchange market, J.~Econ. Interact. Coord. {\bf 4}, 55-72 (2009).
\bibitem{smith2009} R.D.~Smith, The spread of the credit crisis: view from a stock correlation network, J.~Kor. Phys. Soc. {\bf 54}, 2460-2463 (2009).
\bibitem{eryigit2009} M.~Eryigit, R.~Eryigit, Network structure of cross-correlations among the world market indices, Physica A {\bf 388}, 3551-3562 (2009).
\bibitem{eom2009} C.~Eom, G.~Oh, H.~Jeong, S.~Kim, Topological properties of stock networks based on minimal spanning tree and random matrix theory in financial time series, Physica A {\bf 388}, 900-906 (2009).
\bibitem{aste2010} T.~Aste, W.~Shaw, T.~Di~Matteo, Correlation structure and dynamics in volatile markets, New J.~Phys. {\bf 12}, 085009 (2010).
\bibitem{keskin2011} M.~Keskin, B.~Deviren, Y.~Kocakaplan, Topology of the correlation networks among major currencies using hierarchical structure methods, Physica A {\bf 390}, 719-730 (2011).
\bibitem{zhang2011} Y.~Zhang, G.H.T.~Lee, J.C.~Wong, J.L.~Kok, M.~Prusty, S.A.~Cheong, Will the US economy recover in 2010? A minimal spanning tree study, Physica A {\bf 390}, 2020-2050 (2011).
\bibitem{sandoval2012} L.~Sandoval~Jr., Pruning a minimum spanning tree, Physica A {\bf 391}, 2678-2711 (2012).
\bibitem{kwapien2012} J.~Kwapie\'n, S.~Dro\.zd\.z, Physical approach to complex systems, Phys. Rep. {\bf 515}, 115-127 (2012).
\bibitem{zheng2013} Z.~Zheng, K.~Yamasaki, J.N.~Tenenbaum, H.E.~Stanley, Carbon-dioxide emissions trading and hierarchical structure in worldwide finance and commodities markets, Phys. Rev.~E {\bf 87}, 012814 (2013).
\bibitem{wilinski2013} M.~Wiliński, A. Sienkiewicz, T.~Gubiec, R.~Kutner, Z.R.~Struzik, Structural and topological phase transitions on the German Stock Exchange, Physica A {\bf 392}, 5963-5973 (2013).
\bibitem{wang2013a} G.-J.~Wang, C.~Xie, Y.-J.~Chen, S.~Chen, Statistical properties of the foreign exchange network at different time scales: evidence from detrended cross-correlation coefficient and minimum spanning tree, Entropy {\bf 15}, 1643-1662 (2013).
\bibitem{kantar2014} E.~Kantar, B.~Deviren, M.~Keskin, Hierarchical structure of the European countries based on debts as a percentage of GDP during the 2000-2011 period, Physica A {\bf 414}, 95-107 (2014).
\bibitem{skowron2015} P.~Skowron, M.~Karpiarz, A.~Fronczak, P.~Fronczak, Spanning trees of the world trade web: real-world data and the gravity model of trade, Acta Phys. Pol.~A {\bf 127}, A123-A128 (2015).
\bibitem{kruskal1956} J.~Kruskal, On the shortest spanning subtree of a graph and the traveling salesman problem, Proc. Am. Math. Soc. {\bf 7}, 48-50 (1956).
\bibitem{prim1957} R.C.~Prim, Shortest connection networks and some generalizations, Bell System Techn.~J. {\bf 36}, 1389-1401 (1957).
\bibitem{drozdz2001} S.~Drożdż, F.~Gr\"ummer, F.~Ruf, J.~Speth, Towards identifying the world stock market cross-correlations: DAX versus Dow Jones, Physica A {\bf 294}, 226-234 (2001).
\bibitem{peng1994} C.-K.~Peng, S.V.~Buldyrev, S.~Havlin, M.~Simons, H.E.~Stanley, A.L.~Goldberger, Mosaic organization of DNA nucleotides, Phys. Rev.~E {\bf 49}, 1685-1689 (1994).
\bibitem{podobnik2008} B.~Podobnik, H.E.~Stanley, Detrended cross-correlation analysis: a new method for analyzing two non-stationary time series, Phys. Rev. Lett. {\bf 100}, 084102 (2008).
\bibitem{zebende2011} G.F.~Zebende, DCCA cross-correlation coefficient: quantifying level of cross-correlation, Physica A {\bf 390}, 614-618 (2011).
\bibitem{vassoler2012} R.T.~Vassoler, G.F.~Zebende, DCCA cross-correlation coefficient apply in time series of air temperature and air relative humidity, Phys.~A {\bf 391}, 2438-2443 (2012).
\bibitem{zebende2013} G.F.~Zebende, M.F.~da Silva, A.~Machado Filho, DCCA cross-correlation coefficient differentiation: Theoretical and practical approaches, Phys.~A {\bf 392}, 1756-1761 (2013).
\bibitem{reboredo2014} J.C. Reboredo, M.A. Rivera-Castro, G.F. Zebende, Oil and US dollar exchange rate dependence: a detrended cross-correlation approach, Energy Econ. {\bf 42}, 132-139 (2014).
\bibitem{dasilva2015} M.F.~da~Silva, E.J.A.L.~Pereira, A.M.~da~Silva Filho, A.P.N.~de~Castro, J.G.V.~Miranda, G.F.~Zebende, Quantifying cross-correlation between Ibovespa and Brazilian blue-chips: the DCCA approach, Physica A {\bf 424}, 124-129 (2015).
\bibitem{hussain2017} M.~Hussain, G.F.~Zebende, U.~Bashir, D.~Donghong, Oil price and exchange rate co-movements in Asian countries: Detrended cross-correlation approach, Physica A {\bf 465}, 338-346 (2017).
\bibitem{kwapien2015} J.~Kwapień, P.~Oświęcimka, S.~Drożdż, Detrended fluctuation analysis made flexible to detect range of cross-correlated fluctuations, Phys.~Rev.~E {\bf 92}, 052815 (2015).
\bibitem{kantelhardt2002} J.W.~Kantelhardt, S.A.~Zschiegner, A.~Bunde, S.~Havlin, E.~Koscielny-Bunde, H.E.~Stanley, Multifractal detrended fluctuation analysis of nonstationary time series, Physica A {\bf 316}, 87 (2002).
\bibitem{oswiecimka2014} P.~Oświęcimka, S.~Drożdż, M.~Forczek, S.~Jadach, J.~Kwapień, Detrended cross-correlation analysis consistently extended to multifractality, Phys. Rev.~E {\bf 89}, 023305 (2014).
\bibitem{hosking1981} J.R.M.~Hosking, Fractional differencing, Biometrika {\bf 61}, 165-176 (1981).
\bibitem{tickers} Stock tickers (1999): AA ABT AHP AIG ALD AMGN AOL ARC AT AUD AXP BA BAC BEL BK BLS BMY BUD C CA CBS CCL CCU CHV CL CMB CMCS COX CPQ CSCO DD DELL DH DIS DOW EDS EK EMC EMR ENE F FBF FNM FRE FTU G GCI GE GLW GM GTE GTW HD HWP IBM INTC JNJ JPM KMB KO LLY LOW LU MCD MDT MER MMC MMM MO MOT MRK MSFT MTC MWD ONE ORCL PEP PFE PG PNU QCOM QWST SBC SCH SGP SLB SUNW T TWX TX TXN UMG USW UTX VIAB WCOM WFC WMT XON YHOO (TAQ database, http://www.taq.com).
\bibitem{plerou1999} V.~Plerou, P.~Gopikrishnan, L.A.N.~Amaral, M.~Meyer, H.E.~Stanley, Scaling of the distribution of price fluctuations of individual companies, Phys. Rev.~E {\bf 60}, 6519-6529 (1999).
\bibitem{drozdz2003} S.~Dro\.zd\.z, J.~Kwapie\'n, F.~Gr\"ummer, F.~Ruf, J.~Speth, Are the contemporary financial fluctuations sooner converging to normal?, Acta Phys. Pol.~B {\bf 34}, 4293-4306 (2003).
\bibitem{drozdz2007} S.~Dro\.zd\.z, M.~Forczek, J.~Kwapie\'n, P.~O\'swi\c ecimka, R.~Rak, Stock market return distributions: from past to present, Physica A {\bf 383}, 59-64 (2007).
\bibitem{rak2013} R.~Rak, S.~Drożdż, J.~Kwapień, P.~Oświęcimka, Stock returns versus trading volume: is the correspondence more general?, Acta Phys. Pol.~B {\bf 44}, 2035-2050 (2013).
\bibitem{plerou2002} V.~Plerou, P.~Gopikrishnan, B.~Rosenow, L.A.N.~Amaral, T.~G\"uhr, H.E.~Stanley, Random matrix approach to cross correlations in financial data, Phys. Rev.~E {\bf 65}, 066126 (2002).
\bibitem{kwapien2002} J.~Kwapień, S.~Drożdż, F.~Gr\"ummer, F.~Ruf, J.~Septh, Decomposing the stock market intraday dynamics, Physica A {\bf 309}, 171-182 (2002).
\bibitem{wang2013b} G.-J.~Wang, C.~Xie, S.~Chen, J.-J.~Yang, M.-Y.~Yang, Random matrix theory analysis of cross-correlations in the US stock market: evidence from Pearson’s correlation coefficient and detrended cross-correlation coefficient, Physica A {\bf 392}, 3715-3730 (2013).
\bibitem{epps1979} T.W.~Epps, Comovements in stock prices in the very short run, J.~Am. Stat. Assoc. {\bf 74}, 291-298 (1979).
\bibitem{kwapien2004}  J.~Kwapień, S.~Drożdż, J.~Speth, Time scales involved in emergent market coherence, Physica A {\bf 337}, 231-242 (2004).
\bibitem{toth2005} B.~T\'oth, J.~Kert\'esz, Increasing market efficiency: evolution of cross-correlations of stock returns, Physica A {\bf 360}, 505-515 (2005).
\bibitem{theiler1992} J.~Theiler, S.~Eubank, A.~Longtin, B.~Galdrikian, J.D.~Farmer, Testing for non-linearity in time series: the method of surrogate data, Physica D {\bf 58}, 77-94 (1992).
\bibitem{schuster2006} H.G.~Schuster, W.~Just, Deterministic Chaos: An Introduction, 4th Ed., Wiley (2006).

\end{thebibliography}
\end{document}